\documentclass[twocolumn]{svjour3}         
\smartqed  
\usepackage{graphicx}
\usepackage{amsmath}
\usepackage{multirow}
\usepackage{url}

\journalname{Journal of Computational Electronics}
\begin{document}

\title{Modified valence force field approach for phonon dispersion: from zinc-blende bulk to nanowires 
}
\subtitle{Methodology and computational details}


\author{Abhijeet Paul         \and
        Mathieu Luisier \and 
        Gerhard Klimeck
}

\authorrunning{Paul et. al} 

\institute{   
			  Abhijeet Paul, Mathieu Luisier and Gerhard Klimeck \at
              School of Electrical and Computer Engineering and \\
              Network for Computational Nanotechnology \\
              Purdue University, West Lafayette, USA 47907 \\
              Tel.: 1-765-40-43589 \email{abhijeet.rama@gmail.com}           
}

\date{Received: date / Accepted: date}

\maketitle

\begin{abstract}

The correct estimation of thermal properties of ultra-scaled CMOS and thermoelectric semiconductor devices demands for accurate phonon modeling in such structures. This work provides a detailed description of the modified valence force field (MVFF) method to obtain the phonon dispersion in zinc-blende semiconductors. The model is extended from bulk to nanowires after incorporating proper boundary conditions. The computational demands by the phonon calculation increase rapidly as the wire cross-section size increases. It is shown that the nanowire phonon spectrum differ considerably from the bulk dispersions. This manifests itself in the form of different physical and thermal properties in these wires. We believe that this model and approach will prove beneficial in the understanding of the lattice dynamics in the next generation ultra-scaled semiconductor devices.

\end{abstract}

\section{Introduction}
\label{intro}

The lattice vibration modes known as `phonons' determine many important properties in semiconductors like, (i) the phonon limited low field carrier mobility in mosFETs \cite{JAP_anantram,buin_sinw_phonon}, (ii) the lattice thermal conductivity in semiconductors which plays an important role in thermoelectric design \cite{Mingo_kappa,mingo_ph,thermal_cond_dim}, and (iii) the structural stability of ultra-thin semiconductor nanowires \cite{SINW_110_phonon}. A physics-based method to calculate the phonon dispersion in semiconductors is required to understand and link all these issues. As the device size approach the nanometer scale and as the number of atoms in the structure become countably finite, a continuum material description is no longer accurate. This work provides a complete and elaborate description of an atomistic phonon calculation method based on `Valence Force Field' (VFF) model \cite{Keating_VFF,VFF_mod_herman,VFF_mod_zunger}, a frozen phonon approach, with application to bulk and nanowire structures.

A variety of methods have been reported in the literature for the calculation of the phonon spectrum such as the Valence Force Field (VFF) method and its variants \cite{VFF_mod_herman,VFF_mod_zunger,Keating_VFF,McMurry_VFF}, Bond Charge Model (BCM) \cite{bcm_weber,BCM_model}, Density Functional Method \cite{SINW_110_phonon,jauho_method}, etc. We focus on VFF methods in this work. There are multiple reasons for using a VFF based model: (a) in covalent bonded crystals, like Si, Ge, GaAs, simple VFF potentials are sufficient to match the experimental data \cite{McMurry_VFF}, (b) valence coordinates and hence the potential energy (U) depend only on the relative positions of the atoms and are independent of rigid translations and rotations of the solid, and (c) it is easy to extend the model to confined ultra-scaled structures made of few atoms since the interactions are at the atomic level.

The original Keating VFF model \cite{Keating_VFF} describe the LA, LO and TO phonons reasonably well in zinc-blende materials, however, it does not produce the flatness in the  TA branch in Si, Ge \cite{bcm_weber,VFF_mod_herman}. Also the limitation of the model to correctly describe the elastic constants (C11, C12, C44) in these materials also limits its use \cite{VFF_mod_herman}. In order to extend the available VFF models to nanostructures, we need to identify the models which can correctly describe the phonons in the entire Brillouin zone (BZ) in zinc-blende semiconductors. There are older works where as many as six parameters \cite{six_param_VFF} have been used in VFF to obtain the correct phonon dispersions. However, the task of this work has been to obtain a VFF model which (i) can capture the correct physics and (ii) is computationally not very expensive. Hence such a model can be extended to nanostructures like nanowires, ultra-thin-bodies, etc. To this end we have identified two VFF models which satisy the requirements. The modified VFF (MVFF) model presented here combines these two following models, (i) VFF model from Sui et. al \cite{VFF_mod_herman} which is suitable for non-polar materials like Si and Ge and (ii) VFF model from Zunger et. al \cite{VFF_mod_zunger} which is suitable for treating polar materials like GaP, GaAs, etc. This extended model is called the `MVFF model' in this study.

The main focus of this work is to show the implementation of VFF models for phonon calculation in zinc-blende (diamond) lattices. We present the details on the atomic groups which make up the interactions, the application of boundary conditions in the nanostructures, the eigen value problem, the computational requirements and the evaluation of lattice properties. We benchmarked the model for variety of zinc-blende materials like Si, Ge, GaP, GaAs, etc. In this paper we present the results using Si (sometimes Ge too) as a specific example.  We also present a comparison of the Keating VFF model \cite{Keating_VFF} with the present MVFF model for Si to elucidate the differences in physical results and their computational requirements. 
 
Previous theoretical works have reported the calculation of phonon dispersions in SiNWs using a continuum elastic model and Boltzmann transport equation \cite{continuum_model}, atomistic first principle methods like DFPT (Density Functional Perturbation Theory) \cite{SINW_110_phonon,jauho_method,sinw_cv,strain_effect_1} and atomistic frozen phonon approaches like Keating-VFF (KVFF) \cite{Mingo_kappa,SINW_111}.  Thermal conductivity in SiNWs has been studied previously using the KVFF model \cite{mingo_ph,strain_effect_2}.

This paper has been arranged in the following sections. The MVFF theory (a `frozen phonon' method) for the phonon dispersion in zinc-blende semiconductors is reviewed in Sec. \ref{sec:1}. It provides details about the total potential energy (U) of the crystals in the MVFF model (Sec. \ref{sec:1_1}), construction of the dynamical matrix (DM) (Sec. \ref{sec:1_2}), application of boundary conditions to the DM (Sec. \ref{sec:1_3}), solution of the resulting eigen value problem (Sec. \ref{sec:1_4}), and calculation of sound velocity ($V_{snd}$) (Sec. \ref{sec:1_5}), lattice thermal conductance ($\sigma_{l}$) (Sec. \ref{sec:1_6}) and the mode Gr$\ddot{u}$neisen parameters (Sec. \ref{gparam_section}) using the phonon spectrum. The computational details for the calculation of phonon dispersion are presented in Sec. \ref{sec:2}. Different aspects of the dynamical matrix like the size, fill-factor, sparsity pattern, etc., are provided in Sec. \ref{sec:2_1}. Timing analysis for the assembly of the DM are shown in Sec.\ref{sec:2_2}. Section \ref{sec:3} presents, a benchmark of the MVFF results against experimental data for different semiconductors (Sec. \ref{sec:3_1}), a comparison of the MVFF and Keating-VFF (KVFF) models (Sec.\ref{sec:3_2}), phonon spectrum in Si nanowires (SiNW) with free and clamped boundary conditions (Sec. \ref{sec:3_3}) and lattice thermal conductance using SiNW phonon spectrum (Sec.\ref{sec:3_4}). Conclusions are given in Sec.\ref{sec:4}.

\section{Theory, Approach and Parameters}
\label{sec:1}

In a given system, the phonons are modeled by solving the equations of motion of its atomic vibrations. Since VFF is a crystalline model, the dynamical equation for each atom `i' can be written as, 

\begin{equation}
\label{eq:motion}
m_{i}\frac{\partial^2}{\partial t^2} (\Delta R_{i}) = F_{i} = -\frac{\partial U}{\partial (\Delta R_{i})}
\end{equation} 
where, $\Delta R_{i}$, $F_{i}$ and U are the vibration vector of atom `i', the total force on atom `i' in the crystal, and  the potential energy of the crystal, respectively. Equation (\ref{eq:motion}) indicates that the calculation of the vibrational frequencies requires a good estimation of the potential energy of the system. The next part discusses the calculation of U within the MVFF model.

\subsection{Crystal Potential Energy (U)}
\label{sec:1_1}
The MVFF method \cite{VFF_mod_herman,VFF_mod_zunger,Keating_VFF} approximates the potential energy U, for a zinc-blende (or diamond) crystal, based on the nearby atomic interactions (short-range) \cite{VFF_mod_herman,VFF_mod_zunger} as, 

\begin{figure}[t]
\centering

\includegraphics[width=2.9in,height=1.9in]{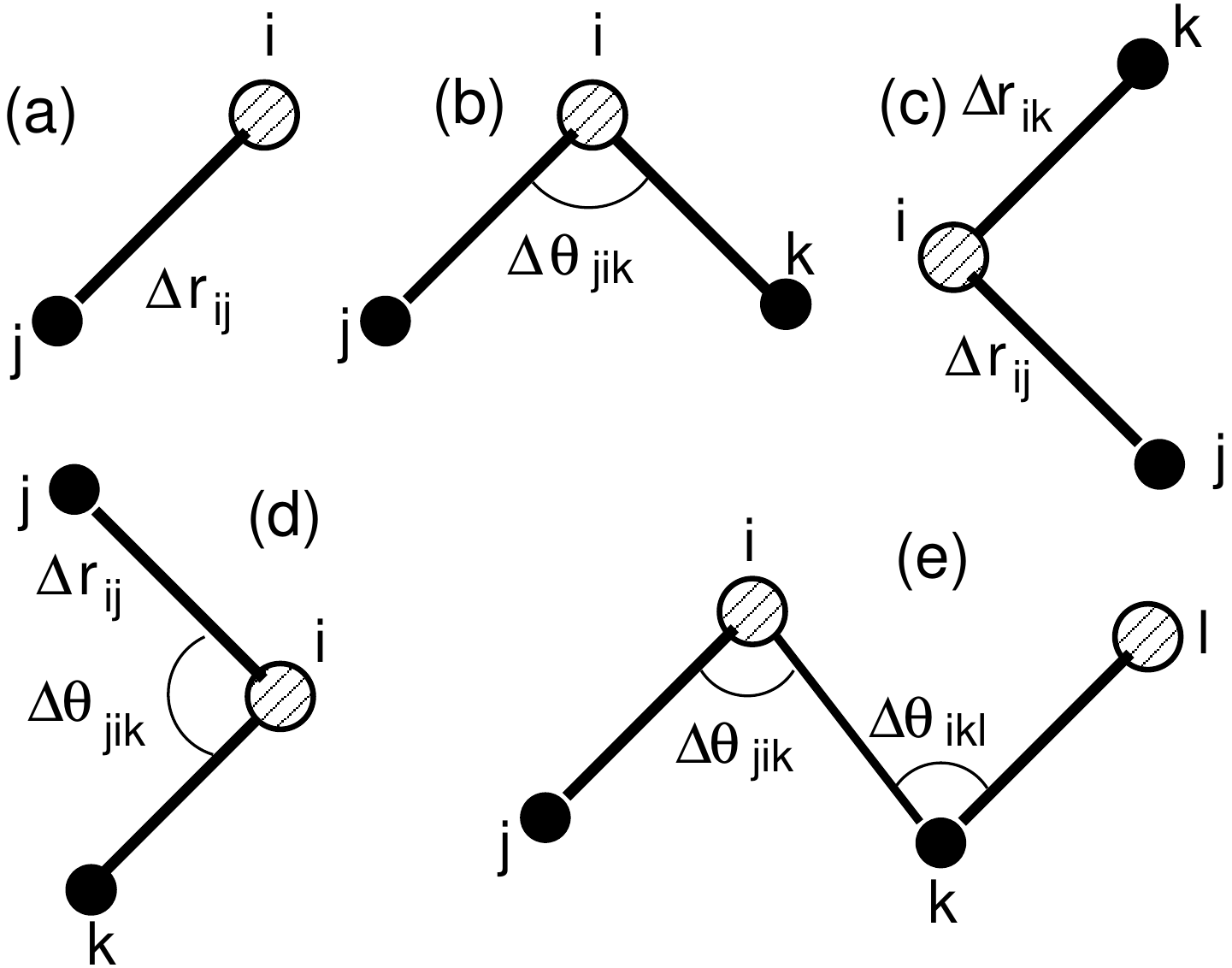} 
\caption{\label{fig1} The short range interactions used for the calculation of phonon dispersion in zinc-blende semiconductors.(a) Bond stretching (b) Bond bending (c) cross bond stretching (d) cross bond bending-stretching and (e) coplanar bond bending interaction.}
\end{figure}

\begin{eqnarray}
\label{eq:uelastic}
 U & \approx & \frac{1}{2} \sum_{i\in N_{A}} \Bigg[ \sum_{j\in nn(i)} U^{ij}_{bs} + \sum^{j \neq k}_{j,k \in nn(i)} \big( U^{jik}_{bb} \nonumber \\
            & &  + U^{jik}_{bs-bs} + U^{jik}_{bs-bb} \big) + \sum^{j \neq k \neq l}_{j,k,l \in COP_{i}} U^{jikl}_{bb-bb} \Bigg]
\end{eqnarray}
where $N_{A}$, $nn(i)$, and $COP_{i}$ represent the total number of atoms in one unitcell, the number of nearest neighbors for atom `i', and the coplanar atom groups for atom `i', respectively. The first two terms in Eq. (\ref{eq:uelastic}) are from the original KVFF model \cite{Keating_VFF}. The other interaction terms are needed to accurately describe the phonon dispersion in the entire Brillioun Zone (BZ). The terms $U^{ij}_{bs}$ and $U^{jik}_{bb}$ represent the elastic energy from bond stretching and bond-bending between the atoms connected to each other (Fig. \ref{fig1}.a,b). The terms $U^{jik}_{bs-bs}$, $U^{jik}_{bs-bb}$ and $U^{jikl}_{bb-bb}$ represent the cross bond stretching \cite{VFF_mod_herman,VFF_mod_zunger}, cross bond bending-stretching \cite{VFF_mod_zunger}, and coplanar bond bending \cite{VFF_mod_herman} interactions, respectively~(Fig. \ref{fig1}. c,d,e). The functional dependence of each interaction term on the atomic positions is given by,

\begin{eqnarray}
\label{u_bs}
& U^{ij}_{bs} &  =  \frac{3} {8}\alpha_{ij}\frac{(r^2_{ij}-d^2_{ij,0})^2} {\|d^2_{ij,0}\|} \\
\label{u_bb}
& U^{jik}_{bb} & = \frac{3}{8}\beta_{jik}\frac{(\Delta \theta_{jik})^2} {\|d_{ij,0}\|\|d_{ik,0}\|} \\
\label{u_bs_bs}
& U^{jik}_{bs-bs} &  = \frac{3}{8}\delta_{jik}\frac{(r^2_{ij}-d^2_{ij,0})(r^2_{ik}-d^2_{ik,0})}{\|d_{ij,0}\|\|d_{ik,0}\|} \\
\label{u_bs_bb}
& U^{jik}_{bs-bb} & =  \frac{3}{8}\gamma_{jik}\frac{(r^2_{ij}-d^2_{ij,0})(\Delta \theta_{jik})} {\|d_{ij,0}\|\|d_{ik,0}\|} \\
\label{u_bb_bb}
& U^{jikl}_{bb-bb} & = \frac{3}{8}\sqrt{(\nu_{jik}\nu_{ikl})} \frac{(\Delta \theta_{jik})(\Delta \theta_{ikl})}{\sqrt{\|d_{ij,0}\|\|d^2_{ik,0}\|\|d_{kl,0}\|}} 
\end{eqnarray}

where $\Delta \theta_{jik} = r_{ij}\cdot r_{ik}-d_{ij,0}\cdot d_{ik,0}$, is the angle deviation of the bond  between atom `i' and 'j' and bond between atom `i' and `k'. The term $r_{ij}$ ($d_{ij,0}$) is the non-ideal (ideal) bond vector from atom `i' to `j'. The coefficients $\alpha$, $\beta$, $\delta$, $\gamma$, and $\nu$ determine the strength of the interactions used in the MVFF model (like spring constants). They are used as fitting parameters to reproduce the bulk phonon dispersion \cite{VFF_mod_herman,VFF_mod_zunger}. The unit of these fitting parameters are in force per unit length (like $Nm^{-1}$). The value of these strength parameters also changes according to the deviation of the bond length and bond angle from their ideal values. This enables the inclusion of the anharmonic properties of the lattice vibrations \cite{anharmonic} in this model. Hence, MVFF is sometimes referred to as `quasi-anharmonic' model. 

\begin{figure}[t]
\centering
\includegraphics[width=2.4in,height=2.2in]{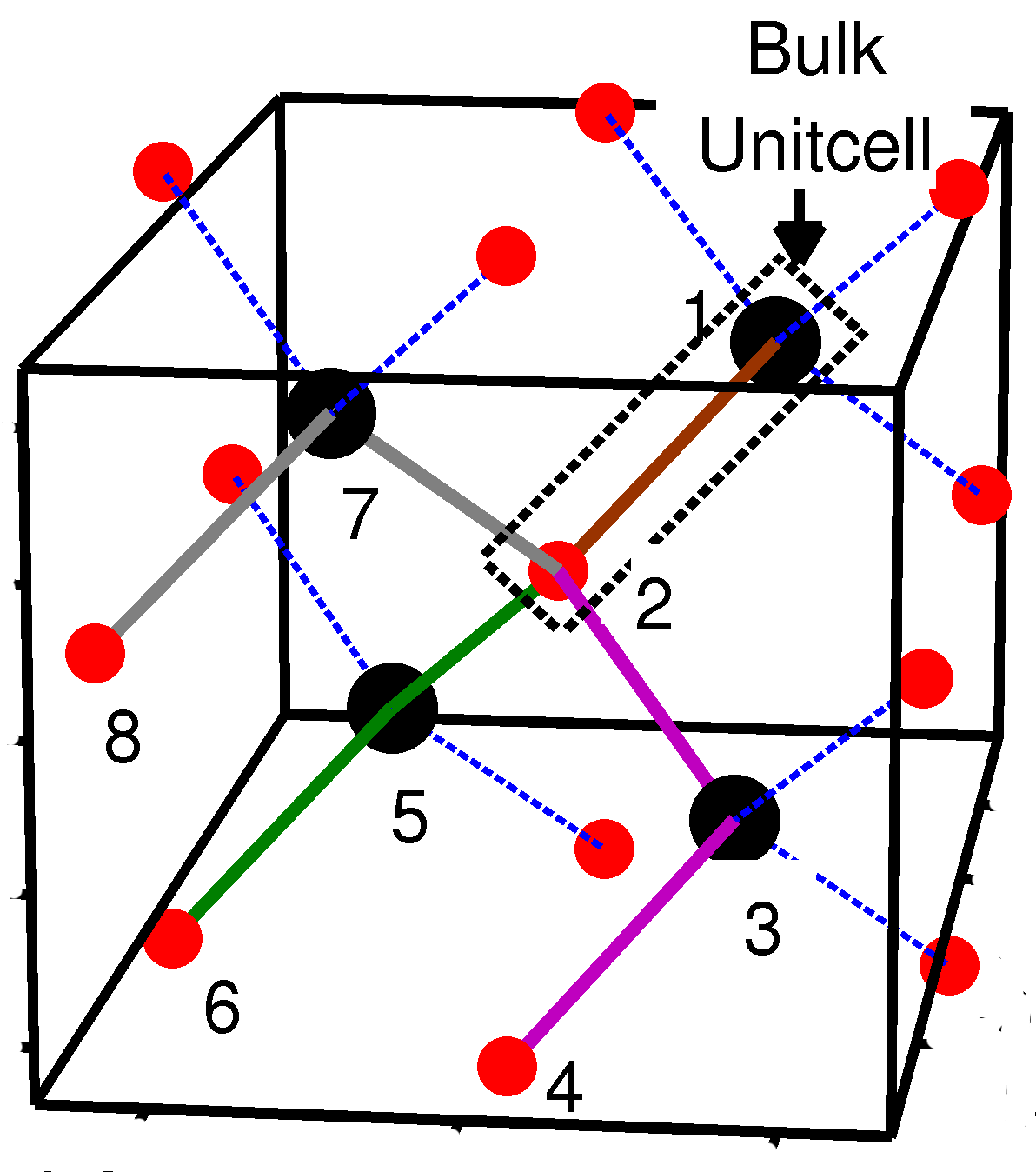} 
\caption{\label{fig:2}(Color online) Three co-planar atom groups (out of 21) shown in a bulk zinc-blende unitcell. The groups are (i) 1-2-3-4, (ii) 1-2-5-6 and (iii) 1-2-7-8. Atoms 1 and 2 form the bulk unitcell used in the calculations. Red (black) atoms are cations (anions).  }
\end{figure}

\textit{Interaction Terms}:~The primitive bulk unitcell used for phonon calculation is made of two atoms (anion-cation pair for zinc-blende and 2 similar atoms for diamond). The black dotted box with atom 1 and 2 represents the bulk primitive unitcell in Fig. \ref{fig:2}. The total number of terms in each interaction in Eq. (\ref{u_bs}-\ref{u_bb_bb}) for a bulk unitcell are provided in Table \ref{tab:1}. Apart from the coplanar bond bending interaction \cite{VFF_mod_herman} all the other terms involve nearest neighbor interactions . There are 21 coplanar (COP) groups  present in a bulk zinc-blende unitcell which are needed for the calculation of the phonon dispersion. For clarity some of these COP groups shown using the number combinations in the caption of Fig. \ref{fig:2}. Each group consists of 4 atom arranged as anion(A)-cation(C)-anion(A)-cation(C) (eg. 1(A)-2(C)-3(A)-4(C) in Fig. \ref{fig:2}). Details about the coplanar interaction groups are provided in Appendix \ref{app:bulk_int}.

\begin{table}[b]
\centering
\caption{Number of terms in different interactions of the MVFF model in a bulk zinc-blende unitcell (anion-cation pair)}
\label{tab:1}       
\begin{tabular}{|l|c|}
\hline
Interaction Type &  Total terms (anion+cation)  \\ 
\hline
Bond stretching (bs) & 8 \\
Bond bending (bb) & 12 \\
Cross bond stretching (bs-bs) & 12 \\
Cross bond stretch-bend (bs-bb) & 12 \\
Coplanar bond bending (bb-bb) & 21 \\
\hline
\end{tabular}
\end{table}

\subsection{Dynamical Matrix (DM)}
\label{sec:1_2}
The dynamical matrix captures the motion of the atoms under small restoring force in a given system.
In this Section we discuss the structure of this matrix.
The derivation of the DM from the equation of motion is given in Appendix \ref{app:dynmat_derive}. The DM calculation is based on the harmonic approximation (see Appendix \ref{app:dynmat_derive}). For the interaction between two atoms `i' and `j', the DM component at atom `i' is given by, 

\begin{equation}
\centering
\label{eq:def_of_dm}
D(ij) = 
\begin{bmatrix}
D^{ij}_{xx} & D^{ij}_{xy} & D^{ij}_{xz}\\
D^{ij}_{yx} & D^{ij}_{yy} & D^{ij}_{yz}\\
D^{ij}_{zx} & D^{ij}_{zy} & D^{ij}_{zz}
\end{bmatrix}
\end{equation}
The 9 components of $D(ij)$ are defined as,

\begin{eqnarray}
\label{eq:Dij_def}
D^{ij}_{mn} & = & \frac{\partial^2 U_{elastic}}{\partial r^{i}_{m}  \partial r^{j}_{n}}, \\ 
			&  & i,j\in N_{A} \quad \text{and} \quad m,n \in [x,y,z], \nonumber
\end{eqnarray} 
where $N_{A}$ is the total number of atoms in the unitcell. For each atom the size of $D(ij)$ is fixed to 3 $\times$ 3. For $N_{A}$ atoms in the unitcell the size of the dynamical matrix is $3N_{A}\times3N_{A}$. However, the matrix is mostly sparse. The sparsity pattern, fill factor, and other related properties of the DM are discussed in Section \ref{sec:2}.

\textit{Symmetry considerations in the DM}: Under the harmonic approximation the dynamical matrix exhibit symmetry properties that can be readily utilized to reduce its assembly time. From software development point of view this is crucial in optimizing matrix construction time, storage and compute times. Due to the continuous nature of the potential energy U, we have, 
 
\begin{equation}
\label{eq:symm_DM}
D^{ij}_{mn} = \frac{\partial^2 U_{elastic}}{\partial r^{i}_{m}  \partial r^{j}_{n}} = \frac{\partial^2 U_{elastic}}{\partial r^{j}_{n}  \partial r^{i}_{m}} = D^{ji}_{nm} 			
\end{equation}
A closer look at Eq. (\ref{eq:symm_DM}) shows the following symmetry relation,

\begin{equation}
\label{eq:dm_symm}
D(ij) = D(ji)' \quad \forall i \neq j .
\end{equation}

This reduces the total number of calculations required to construct the dynamical matrix and speeds up the calculations. Also if the matrix is stored for repetitive use, then only one of the symmetry blocks needs to be stored. This reduces the memory requirement in the software by a factor of 2. Further reduction in the construction time of DM can be achieved depending on the type of interaction, the symmetry of the crystal, and some implementation tricks (not covered in this work, see Ref.\cite{DM_element_reduction} for more discussion). Also the knowledge about the underlying symmetry of the matrix can also help in the selection of linear algebra approaches which can reduce the final solution time (not covered in this work).

\begin{table}[b!]
\centering
\caption{Boundary conditions (BC) in DM based on the dimensionality of the structure}
\label{tab:2}       
\begin{tabular}{|l|c|c|}
\hline
Dimensionality & Periodic BC & Finite Edge BC  \\
\hline
Bulk (3D) & 3 & 0 \\
Thin Film (2D) & 2 & 1 \\
Wire (1D) & 1 & 2 \\
Quantum Dot (0D) & 0 & 3 \\
\hline
\end{tabular}
\end{table}
\subsection{Boundary conditions (BC)}
\label{sec:1_3}
To calculate the eigenmodes of the lattice vibration, it is important to apply appropriate boundary conditions to the DM. In the case of bulk material, the unitcell has periodic (Born-Von Karman) boundary conditions along all the directions (x,y,z) \cite{VFF_mod_herman,bcm_weber} since the material is assumed to have infinite extent in each direction. However, for nanostructures the boundary conditions are different due to the finite extent of the material along certain directions. The boundary conditions vary depending on the dimensionality of the structure (1D, 2D or 3D, see Table \ref{tab:2}) for which the dynamical matrix is constructed. There are 2 types of boundary conditions; (i) Periodic Boundary condition (PBC) which assumes infinite material extent in a particular direction and (ii) Finite Edge boundary conditions (FEBC) such as open or clamped, which assumes finite material extent in a particular direction. Table \ref{tab:2} provides the boundary condition details depending on the dimensionality of the structure used for phonon calculation.

The use of PBC has been discussed in many papers like \cite{VFF_mod_herman,VFF_mod_zunger,bcm_weber}. In this work we consider the boundary conditions associated with geometrically confined nanostructures. The vibrations of the surface atoms can vary from completely free (free BC) to damped oscillations (damped BC). It is shown next that all these cases can be handled within one single boundary condition.  

\begin{figure}[t!]
\centering
\includegraphics[width=2.6in,height=2.4in]{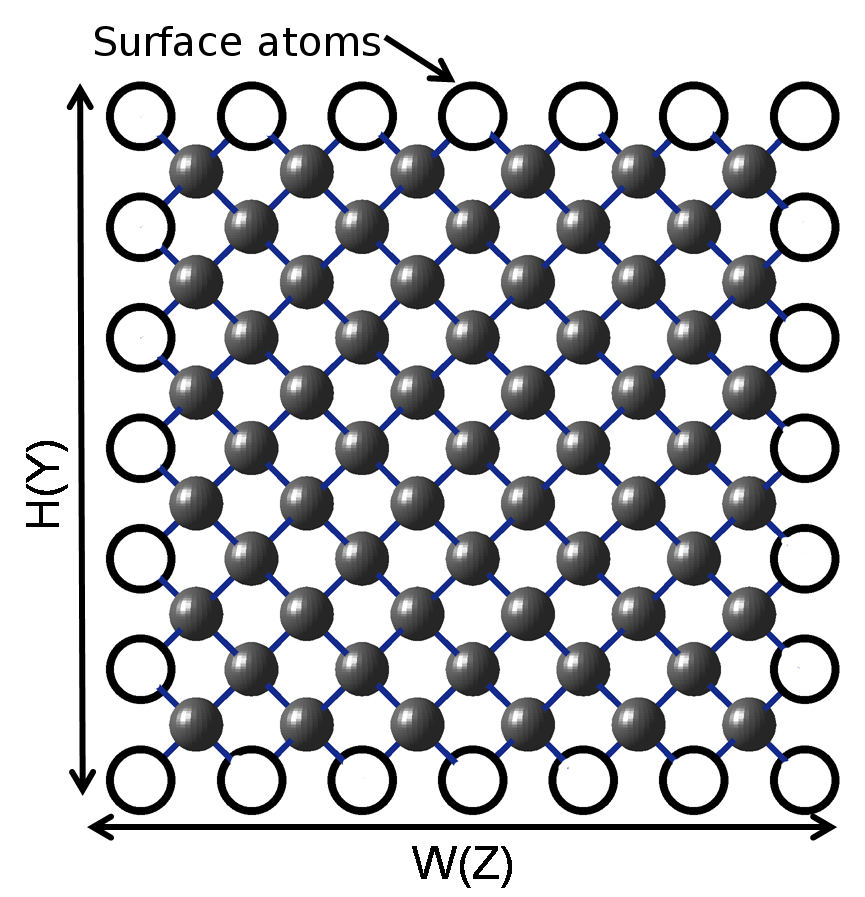} 
\caption{\label{fig:nw_unitcell}Projected unitcell of a $\langle$100$\rangle$ oriented rectangular SiNW shown with surface (hollow) and inner (gray filled) atoms.}
\end{figure}

\textit{Boundary conditions for nanostructures:} The surface atoms (Fig. \ref{fig:nw_unitcell}, hollow atoms) of the nanostructures can vibrate in a very different manner compared to the inner atoms (Fig. \ref{fig:nw_unitcell}, filled atoms) since the surface atoms have different number of neighbors and ambient environment compared to the inner atoms. The degree of freedom of the surface atoms can be represented by a direction dependent damping matrix $\Xi$, defined in Appendix \ref{app:A}. In such a case the dynamical matrix component between atom `i' and `j' ($D(ij)$) is modified to,

\begin{equation}
\label{eq:new_dij}
\tilde{D}(ij) = \Xi^{i} D(ij) \Xi^{j}
\end{equation}


\subsection{Diagonalization of the dynamical matrix}
\label{sec:1_4}
After setting up the dynamical matrix with appropriate BCs the following eigen-value problem must be solved, 
\begin{equation}
\label{eq:DM_eig}
D Q(\lambda,q) =  M\omega^{2}(\lambda,q)Q(\lambda,q),
\end{equation}
where, M is the atomic mass matrix. $\lambda$ and $q$ are the phonon polarization and momentum vector respectively. The term $Q(\lambda,q)$ is a column vector containing all the phonon eigen displacement modes $u(\lambda,q)$ associated with the polarization $\lambda$ and momentum $q$. For simplified numerical calculation slightly modifying Eq. (\ref{eq:DM_eig}) leads to,

\begin{equation}
\label{eq:final_DM_eig}
\overline{D} Q(\lambda,q) =  \omega^{2}(\lambda,q) Q(\lambda,q)
\end{equation}
The detail for obtaining $\overline{D}$ is outlined in Appendix \ref{app:B}. To obtain Eq. (\ref{eq:final_DM_eig}) another step is needed. The time dependent vibration of each atom ($\Delta R (t)$) are represented as the linear combination of phonon eigen modes of vibration $u(\lambda,q)$ (a complete basis set) as,

\begin{equation}
\label{eq:vib_mode}
\Delta R_{i}(t) = \sum_{q,P} u_{P}(\lambda,q) e^{i(q\cdot R_{i}-\omega t)}
\end{equation} 
where, $P$ is the size of the basis set and $\omega$ the vibration frequency of the modes. Using the result of Eq. (\ref{eq:vib_mode}) in the LHS of  Eq. (\ref{eq:motion}) yields,

\begin{equation}
\label{eq:delR_relation}
m_{i} \frac{\partial^2}{\partial t^2}\Delta R_{i}(t) = -\omega ^2 \sum_{q,P} u_{P}(\lambda,q) e^{i(q\cdot R_{i}-\omega t)}
\end{equation}
After some mathematical manipulations and using Eq. (\ref{eq:DM_eig}) we obtain the final eigen value problem given in Eq. (\ref{eq:final_DM_eig}).

\subsection{Sound Velocity ($V_{snd}$)}
\label{sec:1_5}
A wealth of information can be extracted from the phonon spectrum of solids. One important parameter is the group velocity ($V_{grp}$) of the acoustic branches of the phonon dispersion which gives the velocity of sound ($V_{snd}$) in the solid. Depending on the acoustic phonon branch used for the calculation of $V_{grp}$, the sound velocity can be either, (a) longitudinal ($V_{snd,l}$) or (b) transverse ($V_{snd,t}$). In solids, $V_{snd}$ is obtained near the BZ center (for q $\rightarrow 0$) where $ \omega \sim q $ . Thus, $V_{snd}$ is given by, 
\begin{equation}
V_{snd} = \frac{\partial \omega(\lambda,q)}{\partial q} \Big \vert_{q\rightarrow 0},
\end{equation} 
where $\lambda$ is the either the transverse or longitudinal polarization of the phonon frequency.

\subsection{Thermal Conductance ($\sigma_{l}$)} 
\label{sec:1_6}
Another important physical property of the semiconductors that can be extracted from the phonon spectrum is the lattice thermal conductivity. For a small temperature gradient ($\Delta T$) at the two ends of a semiconductor, the $\sigma_{l}$ is obtained using Landauer approach \cite{Land} as outlined in Refs. \cite{Mingo_kappa,mingo_ph,jauho_method}. For 1D nanowires $\sigma_{l}$ at a temperature `T' is given by \cite{mingo_ph,Wallace},
\begin{equation}
\label{eq:kbal_1D}
\sigma_{l,1D} = \frac{1}{2\pi} \cdot \int^{\omega_{fin}}_{0} \Pi(\omega) \frac{\partial}{\partial T}\Big[\frac{1}{e^{\hbar \omega/k_{B}T}-1}\Big] \hbar \omega d\omega,
\end{equation}
where $\Pi(w)$ is the transmission of a phonon branch at frequency $\omega$, $\hbar$ and $k_{B}$, the reduced Planck's constant and Boltzmann constants, respectively. Equation (\ref{eq:kbal_1D}) is of general validity and involves a low temperature approximation. Scattering causes the conductance to vary with the nanowire’s length (L). In the case of ballistic thermal transport the transmission ($\Pi(\omega)$) is always 1 for all the eigen frequencies. 

\subsection{Mode Gr$\ddot{u}$neisen Parameter ($\gamma_{i}$)}
\label{gparam_section}
One of the advantages of using the MVFF model is its ability to keep track of the phonon frequency shift under crystal stress. Under the action of hydrostatic strain the crystal is compressed without changing its symmetry. With pressure (P) the phonon frequency shifts, which is measured by a unitless parameter called the mode `Gr$\ddot{u}$neisen Parameter' given as

\begin{eqnarray}
\gamma_{i,q} & = &-\frac{\partial(ln(\omega_{i,q})}{\partial ln(V)}, \\
	     & = & \frac{B}{\omega_{i,q}}\cdot \frac{\partial \omega_{i,q}}{\partial P},	
\end{eqnarray}
where, $\omega_{i,q}$ is the eigen frequency for the ith branch at $q$ momentum point. The terms B, P and V are the volume compressibility factor, pressure on the system, and volume of the crystal, respectively. Theoretically this parameter is extracted by calculating the eigen frequencies at ambient condition (P = 0) and at small hydrostatic pressure ($\epsilon = \pm0.02$) and then taking the difference in the calculated frequencies. The modification of the force constants under hydrostatic pressure is outlined in Ref. \cite{VFF_mod_herman}. The value of this parameter at high symmetry points ($\Gamma$, X, etc.) in the BZ can be measured experimentally by Raman scattering spectroscopy \cite{gparam_1}. 


In the remaining Sections we provide the computational details, show results on phonon dispersion in bulk and nanowires and give some results on $V_{snd}$, $\gamma_{i}$ and ballistic $\sigma_{l}$ in semiconductor nanowires.

\begin{figure}[t!]
\centering
\includegraphics[width=3in,height=2.4in]{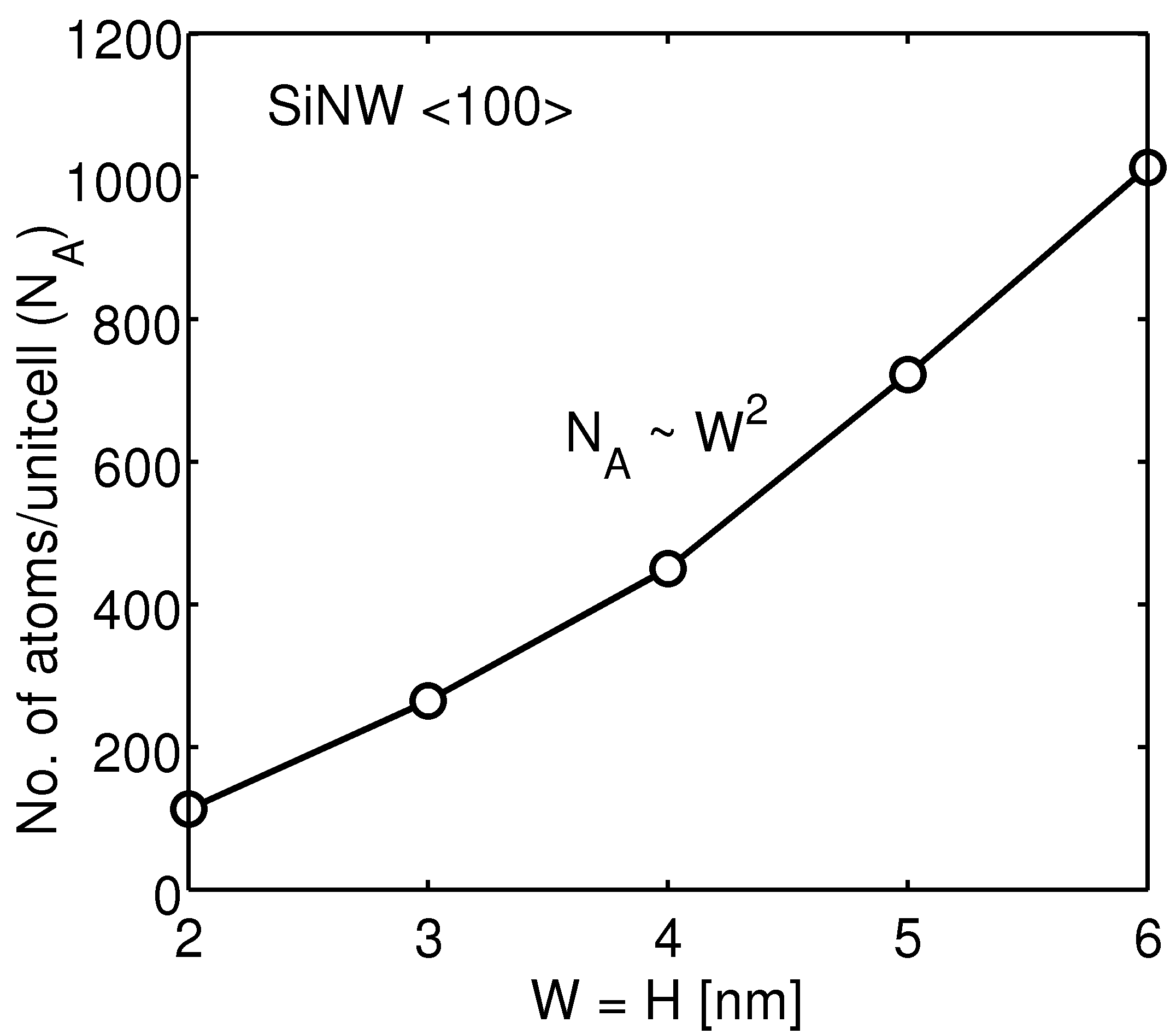} 
\caption{Number of atoms per unitcell ($N_{A}$) with width (W) of $\langle$100$\rangle$ oriented square SiNW.}
\label{fig:na_per_unitcell}
\end{figure}

\begin{figure}[b!]
\centering
\includegraphics[width=3.2in,height=2.1in]{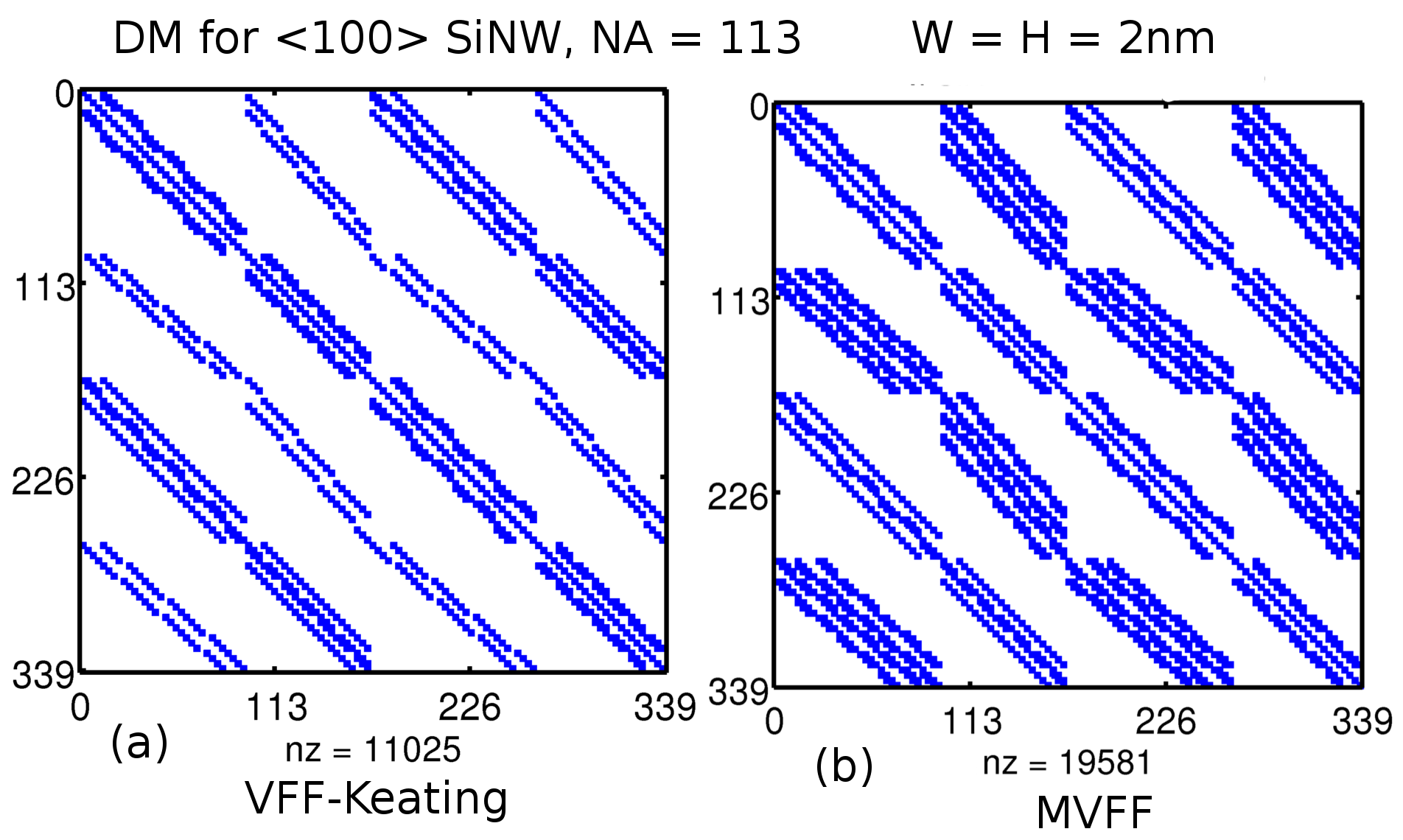} 
\caption{\label{fig:DM_sparsity} Sparsity pattern of the dynamical matrix used in (a) Keating VFF model and (b) MVFF model.  SiNW has W = H = 2nm with 113 atoms in the unitcell.}
\end{figure}


\section{Computational Details}
\label{sec:2}

This Section provides the computational details involved in obtaining the phonon dispersion in semiconductor structures. Details about bulk and nanowire (NW) structures are provided.

\subsection{Dynamical matrix details}
\label{sec:2_1}

A primitive bulk zinc-blende unitcell has 2 atoms. This fixes the size of the DM for the bulk structure to 6 $\times$ 6 ($3N_{A}\times 3N_{A}$) (see Sec.\ref{sec:1_2}). However, for the case of nanowires, $N_{A}$ varies with shape, size and orientation of the wire \cite{jauho_method}. In this paper, all the results are for square shaped SiNW with $\langle$100$\rangle$ orientation. Figure \ref{fig:na_per_unitcell} shows the variation in $N_{A}$ with the width (W) of Silicon NW (SiNW). The number of atoms increase quadratically with W. For a 6nm $\times$ 6nm SiNW, $N_{A}$ is 1013 which means the size of DM is 3,039 $\times$ 3,039. Extrapolating the $N_{A}$ data gives around 7,128 atoms for a 16nm $\times$ 16nm SiNW resulting in a DM of size  21,384 $\times$ 21,384 (details in Appendix \ref{app:DM_prop}). So, the dynamical matrix size increases rapidly with increase in width.

\begin{figure}[t!]
\centering
\includegraphics[width=3.0in,height=2.2in]{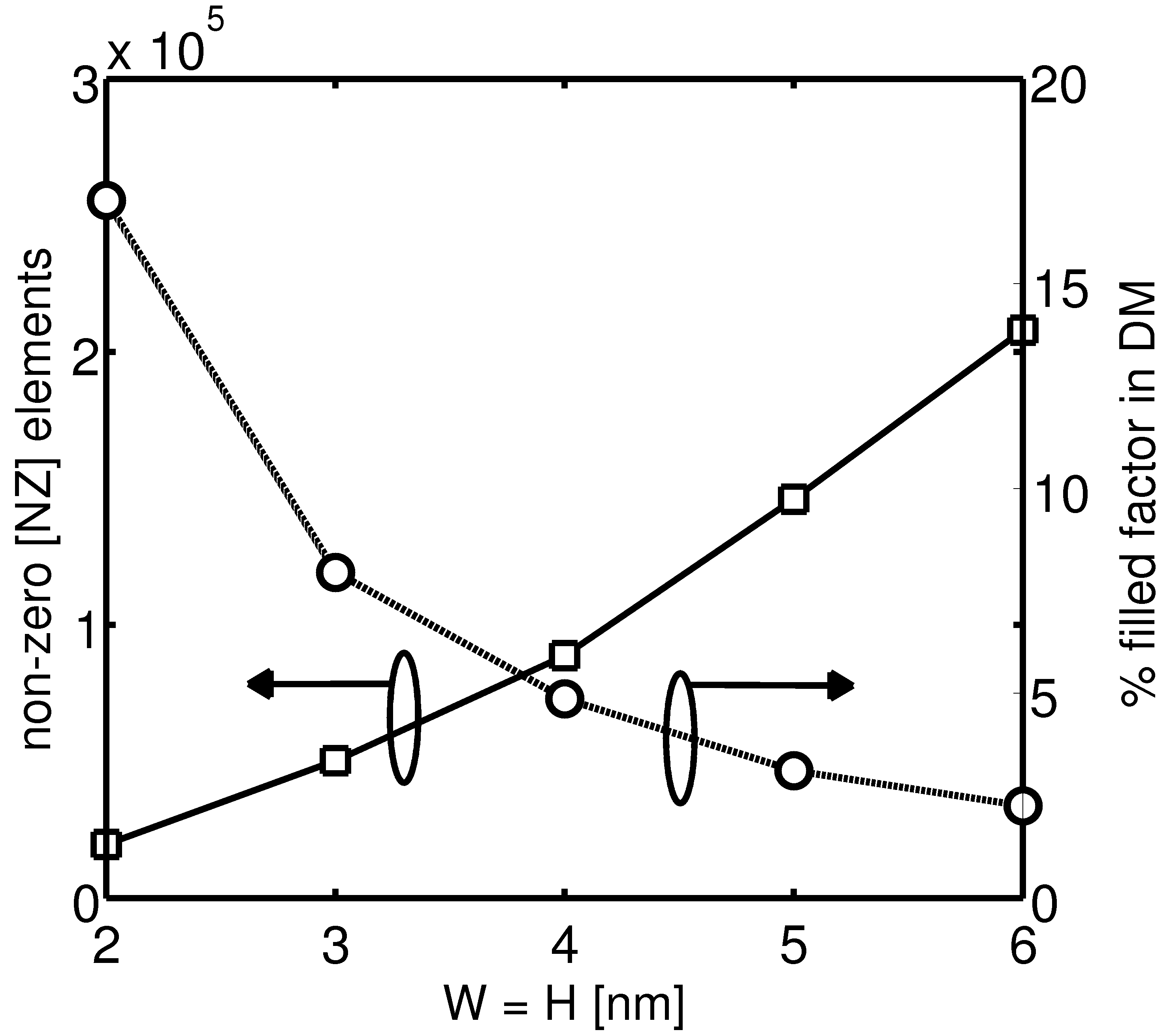} 
\caption{\label{fig:DM_fill_factor}Non zero (NZ) elements in the dynamical matrix and fill factor in DM. Fill factor reduces as the wire unitcell size increases even though the non-zero elements increase.}
\end{figure}

The increase of the DM size with wire cross-section imposes constraint on the structure size which can be solved using the atomistic MVFF method. However, the entire matrix is quite sparse which can be utilized significantly to expand the physical size of the system that can be simulated by using compressed matrix storage methods. The qualitative idea about the filling can be observed from the sparsity pattern for a 2nm $\times$ 2nm SiNW dynamical matrix as shown in Fig. \ref{fig:DM_sparsity}. The quantitative analysis of the fill fraction of the DM and the number of non-zero elements (NZ) in the DM are shown in Fig. \ref{fig:DM_fill_factor}. The non-zero elements in the DM increase quadratically with W of SiNW. An estimate for 16nm $\times$ 16nm SiNW gives about 800,117 non-zero elements (details in Appendix \ref{app:DM_prop}). However, to get an idea about the absolute filling of the DM we define a term called the `fill-factor' given as,
\begin{eqnarray}
fill factor & = & Total\,\,nonzero\,\,elements/Size\,\,of\,\,DM \nonumber \\
		    & = & \frac{NZ}{(3 \times N_{A})^2} \propto \frac{1}{N_{A}} \\
		    && since \;NZ \propto N_{A} \; (see\;Appendix\; Eq.\,(\ref{NZ_with_NA}))\nonumber
\end{eqnarray}
Thus, fill factor varies inversely with number of atoms in the unitcell. The relation of NZ with NA is provided in Appendix \ref{app:DM_prop}.

The percentage fill factor of the DM reduces with increasing W of SiNW (Fig. \ref{fig:DM_fill_factor}). This value is $\sim$ 0.1\% for a SiNW with W $\sim$ 25nm (Appendix \ref{app:DM_prop}) . So even though the non-zero elements increase with W,  DM becomes sparser which allows to store the DM in special compressed format like compressed row/column scheme (CRS/CCS) \cite{CRS_CCS} enabling better memory utilization.

\begin{figure}[t!]
\centering
\includegraphics[width=3.2in,height=1.8in]{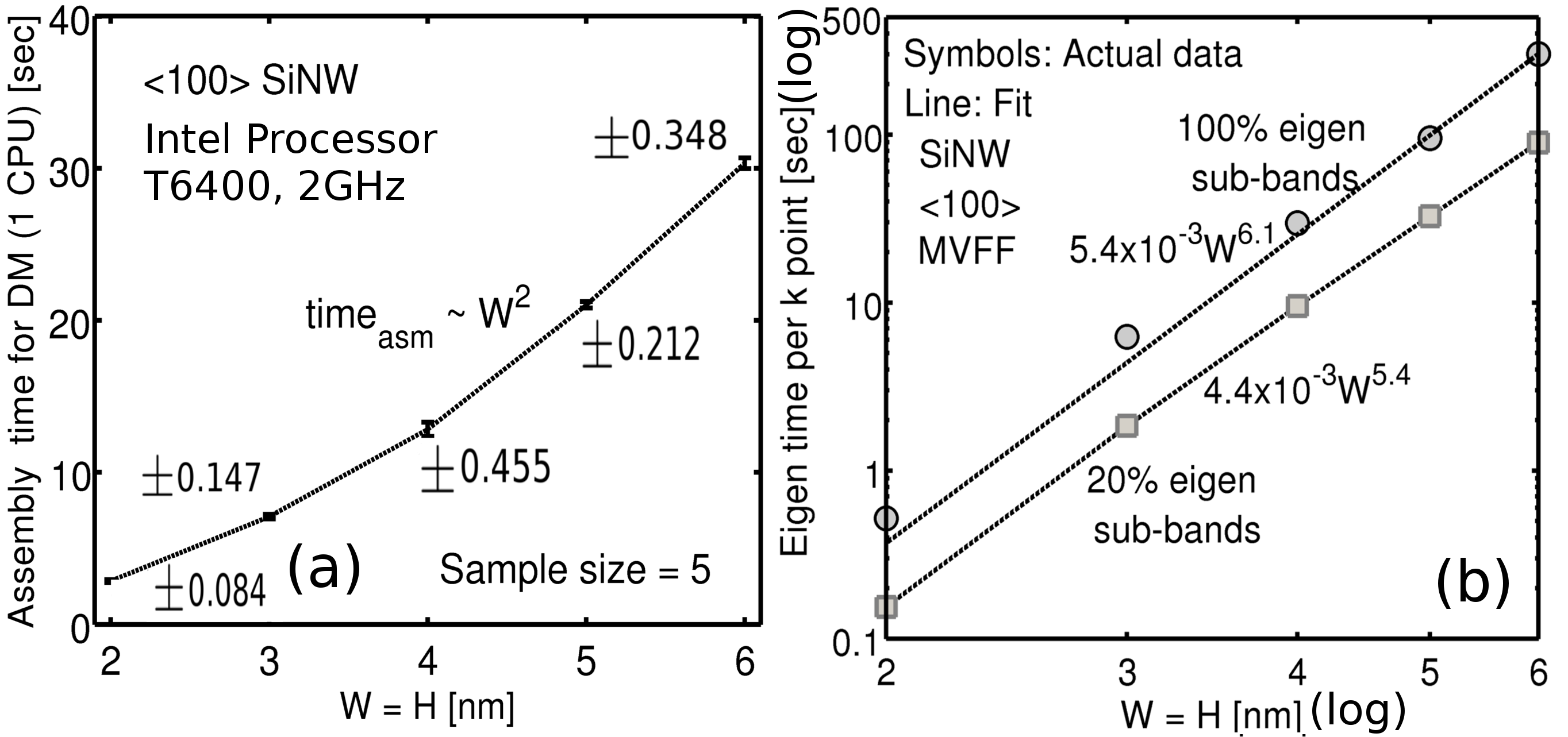} 
\caption{\label{fig:time_asm_dm}(a) Time to assemble the DM ($time_{asm}$) with width (W) of $\langle$100$\rangle$ oriented SiNW. (b) Time to obtain the eigen solutions per k with W for a  $\langle$100$\rangle$ oriented SiNW DM for 100\% and 20\% of the Eigen spectrum. The timing analysis is done on T6400 Intel processor with 2GHz speed. Entire Eigen spectrum along with eigen vectors are obtained using the `eig()' function in MATLAB \cite{eig_matlab}. The partial Eigen spectrum is calculated using the `eigs()' function in MATLAB \cite{eigs_matlab}.}
\end{figure}

\begin{table}[b!]
\centering
\caption{Resource and timing estimate for larger $\langle$100$\rangle$ SiNW}
\label{tab:res_estimate}       
\begin{tabular}{|c|c|c|c|c|c|}
\hline
& & & & & $Time^{*}$  \\
W & $N_{A}$ & NZ  & \% fill & $\overline{time_{asm}}^{*}$ & per k \\
(nm)  &       &     &  factor & (sec)  &  (hours)\\
\hline
16  & 7128  & 800117 & 0.423  & 238.48 & 33.2\\  
20  & 11120 & 1252490 & 0.224 &   370.73 & 129.5\\ 
25  & 17346 & 1.96$\times10^6$ & 0.101 & 576.91 & 505.2 \\
\hline
\multicolumn{6}{l}{$^{*}$\footnotesize{Time estimates on an Intel T6400, 2GHz processor.}}\\
\end{tabular}
\end{table}
\subsection{Timing analysis for the computation of DM}
\label{sec:2_2}

The numerical assembly of the DM takes a considerable time due to the many interactions calculated in the MVFF model.  
The assembly time ($time_{asm}$) increases as $N_{A}$ increases. To give an idea about the timing, the dynamical matrix for SiNW with different W are constructed on a single CPU (Intel T6400, 2GHz processor). The assembly time is calculated for each width 5 times to obtain a mean value for the $time_{asm}$. The error bar at each W is the standard deviation from the mean $time_{asm}$ (Fig. \ref{fig:time_asm_dm}a). In the present case the assembly of the DM is done atom by atom which is useful for distorted materials as well as alloys. The assembly time for the DM in single materials can be reduced dramatically by the assumption of homogeneous bond lengths and a matrix stamping technique \cite{nemo3d}.

After the DM is assembled, it is solved to obtain the eigen modes of oscillations. The time needed to diagonalize ($t_{diag}$) the DM, for each `k' point, using the MATLAB `eig' function \cite{eig_matlab} is also calculated (on the same processor). The $t_{diag}$ value varies as the sixth power of W as shown in Fig. \ref{fig:time_asm_dm}b. However, if only 20\% percent of the Eigen values are calculated the time requirement now goes by the fifth power (shown by the lower line in Fig. \ref{fig:time_asm_dm}b). The Eigen values in this case are calculated using the `eigs' function in matlab \cite{eigs_method}. The calculation of only 20\% of the Eigen spectrum reduces the per-k energy calculation time by $\sim$75\% for a square SiNW with W = H = 6nm. However, the possibility of using only the partial spectrum for the evaluation of the important lattice parameters (like thermal conductance, etc.) is not in the scope of present discussion. For the calculation of physical quantities, in this paper, the complete Eigen spectrum has been used. 

Extrapolating the data for the computational and timing requirement obtained for the smaller SiNWs, can provide some estimates about the size and time requirement for larger SiNWs (Table \ref{tab:res_estimate}). Analytical fits for the variation of the size and time parameters with W are provided in Appendix \ref{app:DM_prop}. 
The timing requirements also help us in the resource requirement estimate for a future Bandstructure Lab extension for phonons on nanoHUB.org \cite{bslab}.

\begin{table}[t!]
\centering
\caption{Force constants (N/m) used for phonon dispersion calculation}
\label{tab:VFF_param}       
\begin{tabular}{|l|c|c|c|c|c|c|}
\hline
Material & Model & $\alpha$ & $\beta$ & $\delta$ & $\gamma$ & $\nu$  \\
\hline
Si  & MVFF \cite{VFF_mod_herman} & 45.1 & 4.89 & 1.36 & 0 & 9.14 \\ 
Si & KVFF \cite{Keating_VFF}& 48.5 & 13.8 & 0 & 0 & 0 \\
Ge & MVFF \cite{VFF_mod_herman}& 37.8 & 4.24 & 0.49 & 0 & 7.62\\  
\hline
\end{tabular}
\end{table}

\begin{figure}[t!]
\centering
\includegraphics[width=3.4in,height=1.85in]{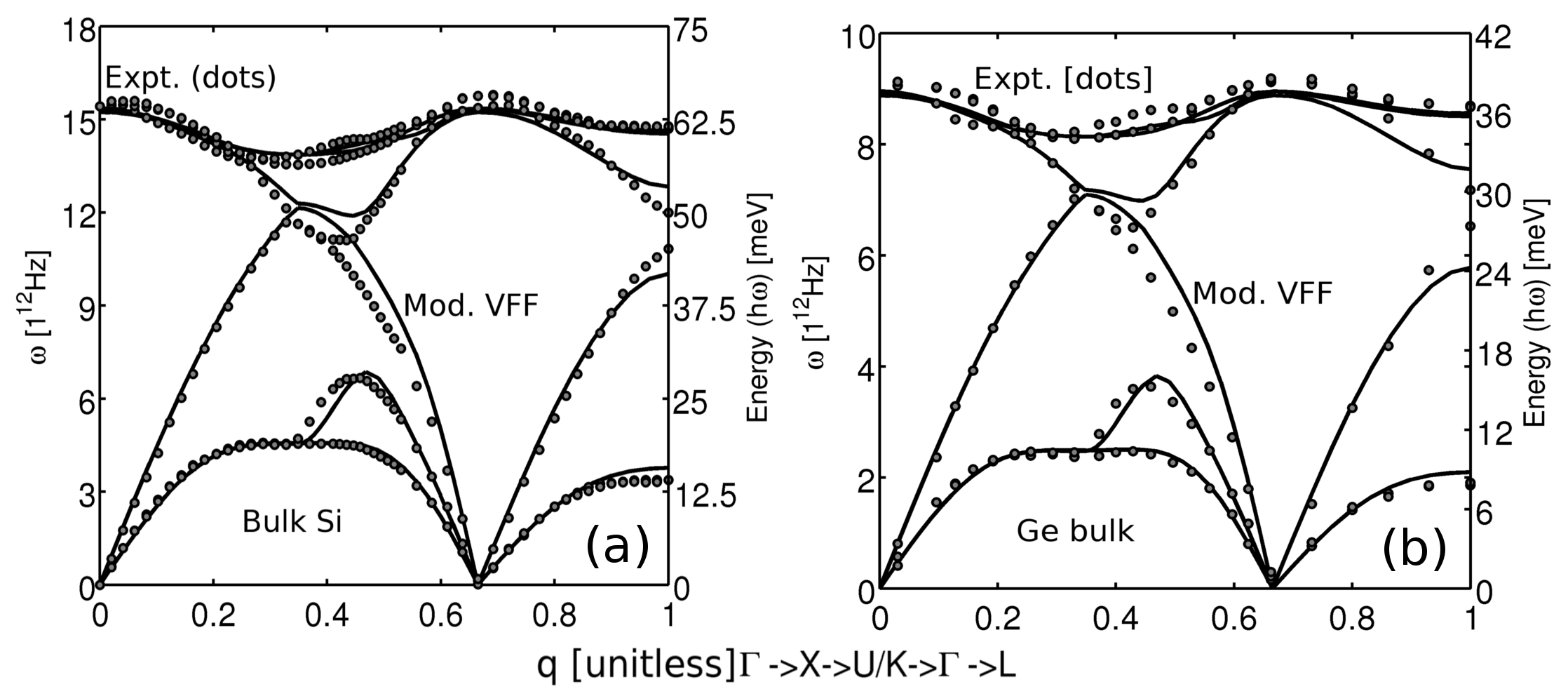} 
\caption{\label{fig:bulk_phonon}Benchmark of simulated bulk phonon dispersion with experimental phonon data for (a) Si and (b) Ge. Experimental data is obtained using neutron scattering at 80K \cite{bulk_si_exp_phon}.}
\end{figure}

\section{Results}
\label{sec:3}
In this Section we show results for the phonon spectrum in bulk and confined semiconductor structures using both the MVFF and KVFF models. Also some of the physical properties extracted from the phonon dispersions are reported.

\subsection{Experimental Benchmarking}
\label{sec:3_1}
The first step to check the correctness of the MVFF model is to compare the simulated results with experimental data. Figure \ref{fig:bulk_phonon} shows the simulated and experimental \cite{bulk_si_exp_phon} bulk phonon dispersion for (a) Silicon and (b) Germanium. The value of the strength parameters are provided in Table \ref{tab:VFF_param}. A very good agreement between the experimental and simulated data is obtained. To further support the correctness of the MVFF model, $V_{snd}$ is calculated in bulk Si and Ge along $\langle$100$\rangle$ direction (Table \ref{tab:vsnd}). The extracted sound velocity compares very well with the experimental sound velocity data \cite{ioffe_online} (max error $\le$ 10\%). 

\begin{table}[b!]
\centering
\caption{Sound Velocity in km/sec in Si, Ge bulk and square nanowires with W = H = 6nm.}
\label{tab:vsnd}       
\begin{tabular}{|c|c|c|c|}
\hline
Material & Structure & $V_{snd}$ calc.  & $V_{snd}$ Expt. \cite{ioffe_online}  \\
\hline
\multirow{4}{*}{Si}  & Bulk $V_{l}$ [100]  & 9.09 & 8.43 ($\sim$8\%) \\
  & Bulk $V_{t}$ [100] & 5.71 & 5.84 ($\sim$2\%) \\
   & NW $V_{l}$ & 6.51 & -- \\
   & NW $V_{t}$ & 4.46 & -- \\
\hline
\multirow{4}{*}{Ge} & Bulk $V_{l}$ [100] & 5.13 & 4.87 ($\sim$5\%) \\
   & Bulk $V_{t}$ [100] & 3.36 & 3.57  ($\sim$6\%)\\
   & NW $V_{l}$  & 3.70 & -- \\
   & NW $V_{t}$  & 2.61 & -- \\
\hline
\end{tabular}
\end{table}

The comparison of second order elastic constants for Si and Ge evaluated using the MVFF model (using the formulation provided in Ref. \cite{VFF_mod_herman}) with experimental data \cite{Madelung} is provided in Table \ref{tab:elastic_param}. The MVFF derived values compare quite well with the experimental data \cite{Madelung}. 

Another comparison carried out to test the robustness of MVFF model is the comparison of the mode Gr$\ddot{u}$neisen parameters at the high symmetry $\Gamma$ and X point with the experimental data (see Table \ref{tab:Gparam_VFF_comparison}). The  MVFF results are in good comparison with the experimental data.

An advantage of using a higher order phonon model is that both phonon dispersions as well the physical parameters can be matched to a good accuracy. Hence, MVFF model captures the experimental phonon dispersion as well as the elastic properties in bulk zinc-blende material very well.

\begin{table}[t!]
\centering
\caption{Elastic constants ($10^{10}Nm^{-2}$) obtained from the MVFF model compared with experimental data \cite{Madelung} for Si and Ge. 
The corresponding errors in the theoretical values are also shown.}
\label{tab:elastic_param}       
\begin{tabular}{|l|l|l|l|l|}
\hline
Material & Model & C11 & C12 & C44  \\
\hline
Si & MVFF  & 16.80 & 6.47  & 7.63   \\ 
Si & Expt. & 16.57 & 6.39 & 7.96  \\ 
Error &  & $\sim$1.4\%    & $\sim$1.2\%  & $\sim$4.14\%  \\ \hline
Ge & MVFF  & 13.22  & 4.84  & 6.29 \\  
Ge & Expt. & 12.40  & 4.13  & 6.83  \\  
Error &  & $\sim$6.6\%    & $\sim$17.2\%  & $\sim$8\%  \\
\hline
\end{tabular}
\end{table}

\subsection{Comparison of VFF models}
\label{sec:3_2}
\begin{figure}[t!]
\centering
\includegraphics[width=3.3in,height=2.1in]{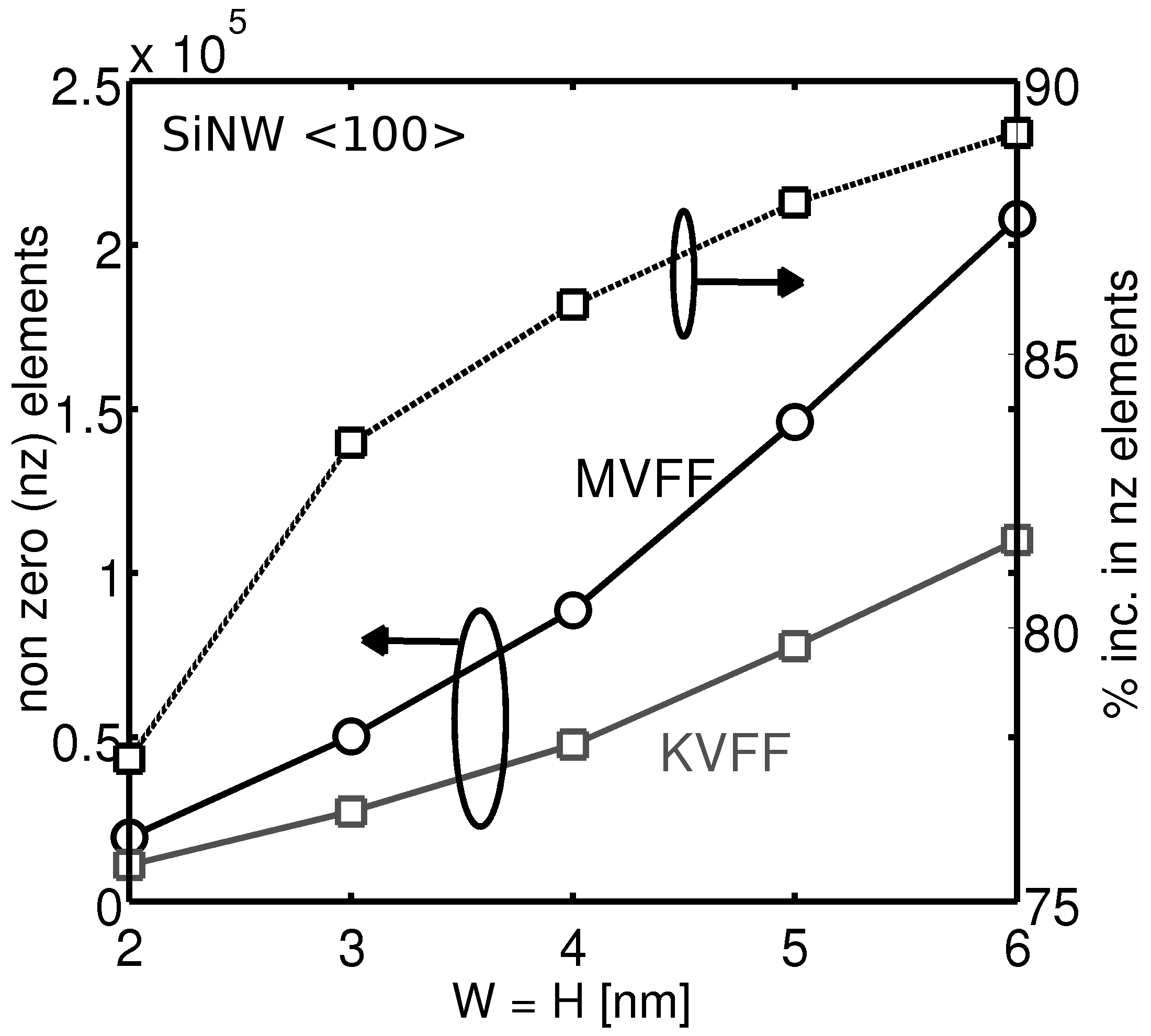} 
\caption{Matrix size and number of non zero elements required by the two models. MVFF has more elements needed for accurate phonon dispersion.}
\label{fig:nz_comparison_model}
\end{figure}

In this Section we compare the original Keating VFF model \cite{Keating_VFF} with the MVFF model to show the need for the more elaborate MVFF model. Both computational requirement and the physical result comparisons are provided in this Section. From computational point of view the DM for both the models are quite different (Fig. \ref{fig:DM_sparsity}). The difference in the sparsity pattern arises because of the coplanar interaction present in the MVFF model which takes into account interactions beyond the nearest neighbors. 
The KVFF model has fewer non-zero elements compared to the MVFF model. The increase in number of NZ elements increases more rapidly in the MVFF model vs. the KVFF model (Fig. \ref{fig:nz_comparison_model}). The MVFF model required twice as many matrix elements compared to the KVFF model for a 5nm $\times$ 5nm SiNW. Thus, MVFF model demands more storage space.

\begin{table}[b!]
\centering
\caption{Comparison of the mode Gr$\ddot{u}$neisen Parameters for bulk Si using the two phonon models.}
\label{tab:Gparam_VFF_comparison}       
\begin{tabular}{|l|c|c|c|c|}
\hline
$\gamma_{i}$ & MVFF & KVFF & Expt./Abinitio & Ref. \\
\hline
$\gamma^{\Gamma}_{LO,TO}$ & 1.05 & 0.81 & $0.98\pm0.06$ & \cite{gparam_1} \\
$\gamma^{\Gamma}_{TA}$ & -0.68 &  -0.43& -0.62 & \cite{VFF_mod_herman}\\
$\gamma^{\Gamma}_{LA}$ & 0.95 & 0.7   & 0.85 & \cite{VFF_mod_herman} \\
$\gamma^{X}_{LO,LA}$ & 1.08 & 0.816 & 1.03 & \cite{gparam_2}\\
$\gamma^{X}_{TO}$ & 1.25 & 0.83 & $1.5\pm0.2$ & \cite{gparam_1} \\
$\gamma^{X}_{TA}$ & -1.58 & -0.33 & $-1.4\pm0.3$& \cite{gparam_1}\\
\hline
\end{tabular}
\end{table}

\begin{figure}[b!]
\centering
\includegraphics[width=3.4in,height=2.0in]{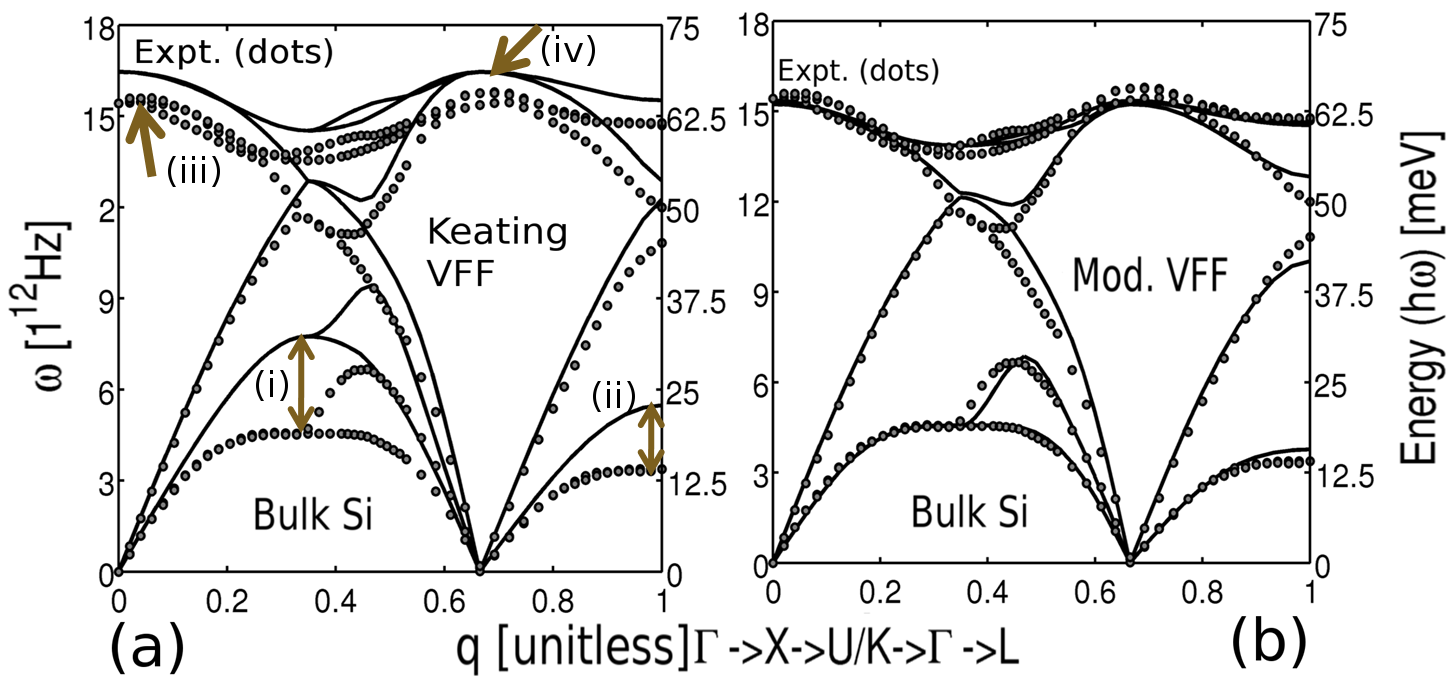} 
\caption{ Comparison of simulated phonon results with experimental data (at 80K from \cite{bulk_si_exp_phon}) from the two phonon models (a) Keating VFF and (b) Modified VFF. The KVFF model fails to reproduce many important features in the experimental data as shown by the arrows in (a). The shortcomings are (i) over estimation of the acoustic mode at X by $\sim$60\%, (ii) acoustic branch over estimated at L by $\sim$95\% and failure to reproduce the correct value for the optical branches altogether pointed in (iii) and (iv).}
\label{fig:bulk_si_phon_comp}
\end{figure}

A comparison of the bulk Si phonon dispersion from the two models is shown in Fig. \ref{fig:bulk_si_phon_comp}. The parameters for both the models are provided in Table \ref{tab:VFF_param}. Qualitatively MVFF shows a better match to the experimental data compared to the KVFF model. There are few important points to be noted in the bulk phonon dispersion. The KVFF model reproduces the acoustic branches very well near the Brillouin zone (BZ) center but overestimates the values near the zone edge (at X and L point in the BZ, Fig. \ref{fig:bulk_si_phon_comp}). The MVFF model overcomes this shortcoming  and reproduces the acoustic branches very well in the entire BZ. Comparison of sound velocity along the $\langle$100$\rangle$ direction for bulk Si obtained from both the models show a very good match to the experimental data (Table \ref{tab:VFF_comparison}).

The KVFF model overestimates the optical phonon branch frequencies whereas the MVFF model produces a very good match to the experimental data (Fig. \ref{fig:bulk_si_phon_comp}) and Table \ref{tab:VFF_comparison}. The comparison of the optical frequency at the $\Gamma$ point reveals that the KVFF model overshoots the experimental value by $\sim$ 7\% whereas the MVFF model is higher by only $\sim$ 0.6\%.

The comparison of the mode Gr$\ddot{u}$neisen parameters for bulk Si using the two models is shown in Table \ref{tab:Gparam_VFF_comparison}. KVFF gives wrong values of these parameters compared to the experimental values. However, MVFF model is able to reproduce the experimental value very well. This shows the importance of using a quasi-harmonic model to correctly obtain the phonon frequency shifts \cite{VFF_mod_herman,gparam_3}. A similar failure of the KVFF model for III-V zinc-blende materials have been reported in Ref. \cite{olga_gparam}. 

The correct representation of bulk phonons is very important since this will affect the phonon spectrum of the confined structures. At the same time the physical properties like lattice thermal conductivity, phonon density of states (DOS), etc. are also affected. Since the MVFF model matches the experimental bulk phonon data more accurately, though at the expense of additional calculations and storage, compared to the original KVFF model, we believe that the MVFF model will give better results for phonon dispersion in nanostructures. 


\subsection{Phonons in nanowires}
\label{sec:3_3}

\begin{figure}[b!]
\centering
\includegraphics[width=3.4in,height=2.2in]{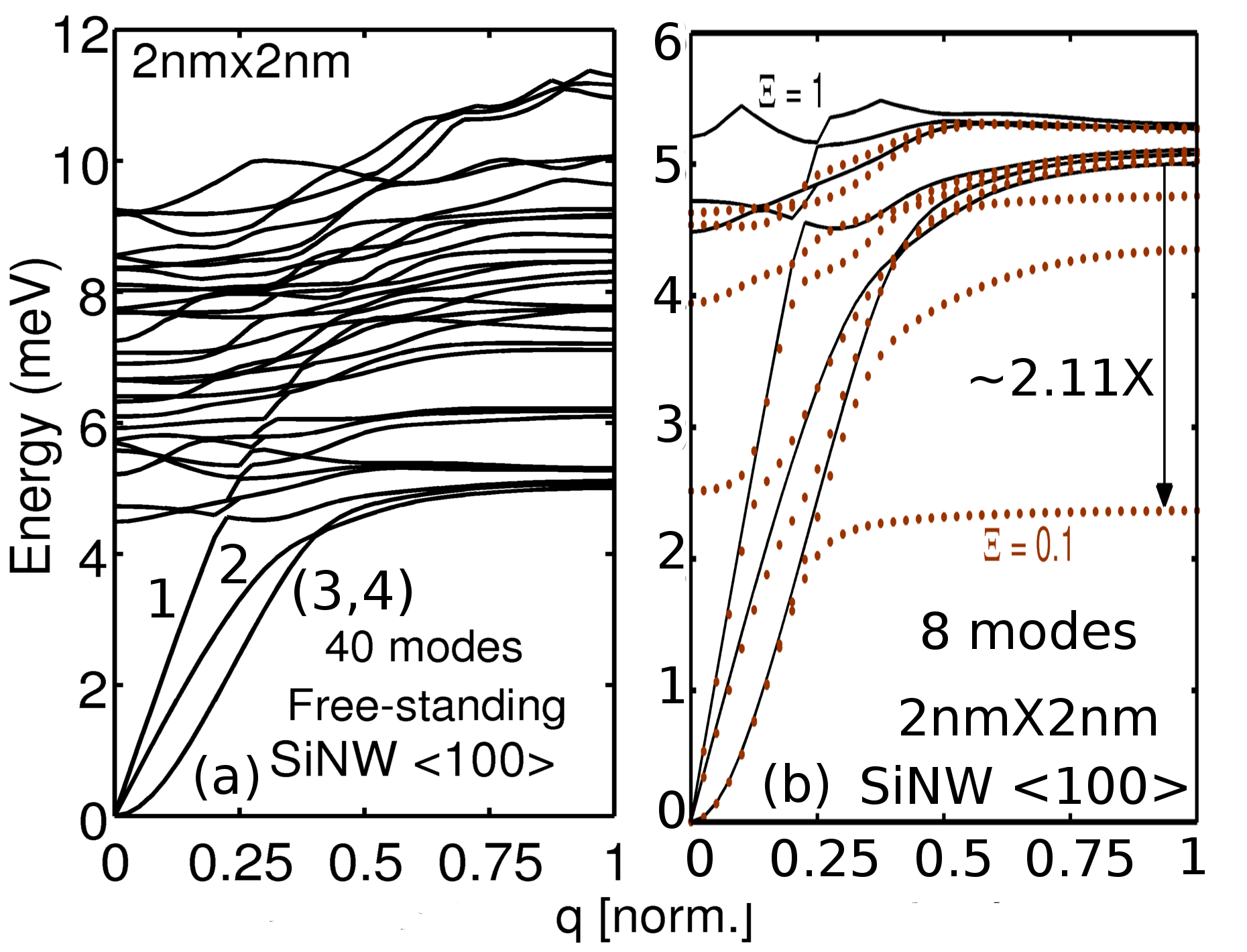} 
\caption{(a) Phonon dispersion in $\langle$100$\rangle$ oriented SiNW with W = H = 2nm. For clarity only the lowest 40 sub-bands are shown. (b) Dependence of phonon dispersion on the damping of vibration of the bottom surface atoms for $\Xi$ = 1 and 0.1. Reduction of phonon energy at the Brillouin zone boundary is stronger compared to the zone center.}
\label{fig:wire_phonon}
\end{figure}

After benchmarking the bulk phonon dispersion, the same parameters are used to obtain the phonon spectrum in $\langle$100$\rangle$ square SiNW (Fig. \ref{fig:wire_phonon}). The result for a 2nm $\times$ 2nm free-standing SiNW is shown in Fig. \ref{fig:wire_phonon}(a). Some of the key features to notice in the phonon dispersion are, (i) presence of two acoustic branches ($\omega(q)\sim $ q, 1,2 in \ref{fig:wire_phonon}(a)), (ii) two degenerate modes (3,4 in \ref{fig:wire_phonon}(a)) with $\omega(q)\sim\,q^2$, which are called the `flexural modes', typically observed in free-standing nanowires \cite{SINW_110_phonon,jauho_method,SINW_111} and (iii) heavy mixing of the higher energy sub-bands leaving no `proper' optical mode. The features are quite different from the bulk phonon spectrum. This will strongly affect other physical properties of nanowires extracted using the phonon dispersion.

\begin{table}[t!]
\centering
\caption{Comparison of bulk parameters in Si for two models}
\label{tab:VFF_comparison}       
\begin{tabular}{|l|c|c|c|}
\hline
\multirow{2}{*}{Model} & $V^{bulk}_{l,100}$ & $V^{bulk}_{t,100}$ & $\omega_{opt}(\Gamma)$ \\
                      & (km/sec) & (km/sec) & (THz)   \\
\hline
MVFF & 9.09 & 5.71 & 15.49  \\
KVFF & 8.35 & 5.75 & 16.46 \\
Expt. & 8.43 \cite{ioffe_online} & 5.84 \cite{ioffe_online} & 15.39 \cite{bulk_si_exp_phon} \\
\hline
\end{tabular}
\end{table}

In Fig. \ref{fig:wire_phonon}b, we explore the effect of a substrate on which the nanowire may be mounted. Only the bottom surface of the SiNW is clamped whereas the other three sides have free boundary condition. Using the boundary condition method discussed in sec.\ref{sec:1_3}, the phonon spectrum in a 2nm $\times$ 2nm $\langle$100$\rangle$ SiNW are calculated with different damping values ($\Xi = 1$ (free standing) and 0.1).  The effect of damping is very prominent at the BZ edge compared to the zone center. Zone edge frequencies decrease in energy as the damping increases. A reduction of $\sim 2.11 \times$ are observed for the zone edge frequency for 1st branch at $\Xi = 0.1$ (Fig. \ref{fig:wire_phonon}(b)) which shows that the NW vibrational energy is decreasing more at higher momentum `q' values. The first four branches are very strongly affected, however, the higher phonon branches are less affected.

\subsection{Ballistic lattice thermal conductance ($\sigma^{bal}_{l}$) in SiNWs}
\label{sec:3_4}
\begin{figure}[t!]
\centering
\includegraphics[width=3.4in,height=1.9in]{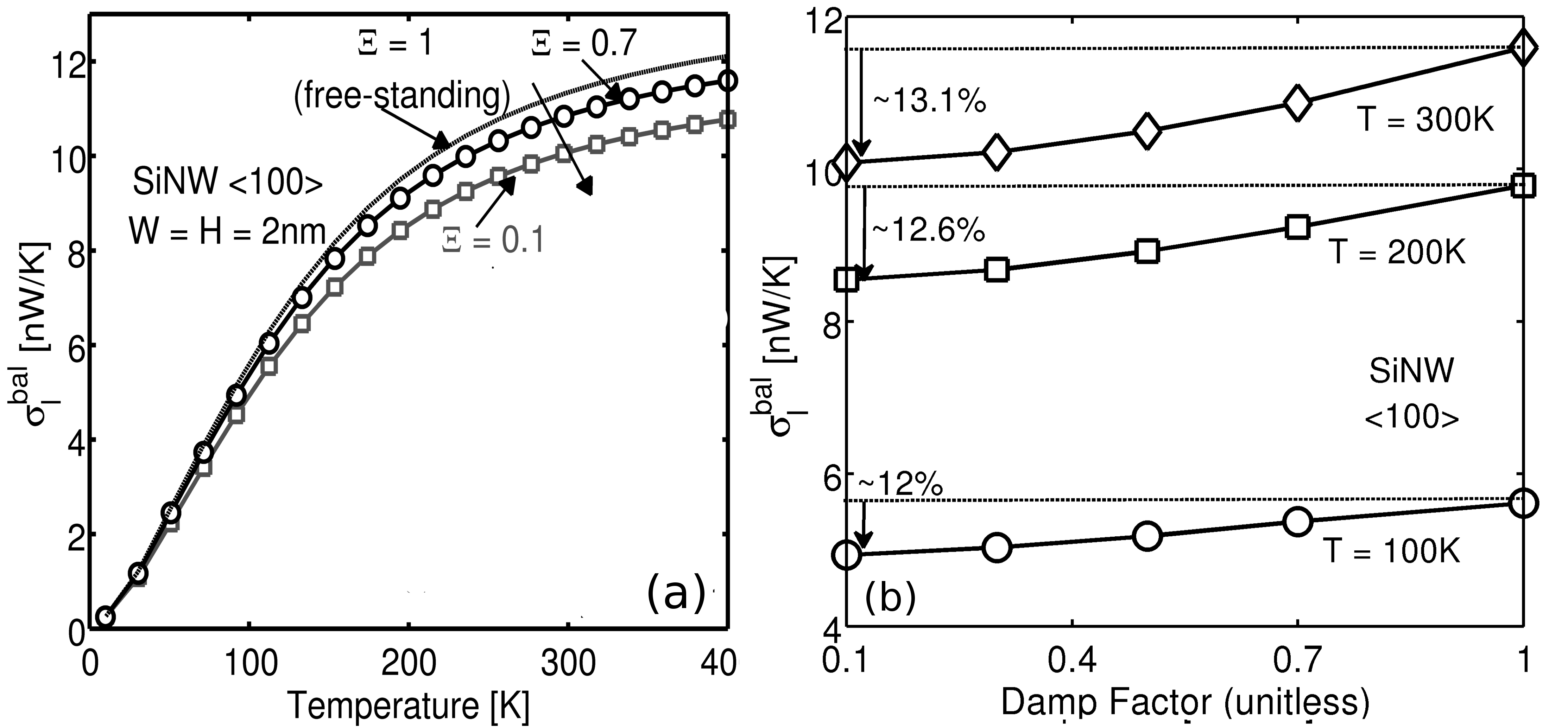} 
\caption{Ballistic lattice thermal conductivity ($\sigma^{bal}_{l}$ for a 2nm $\times$ 2nm $\langle$100$\rangle$ SiNW with different bottom surface damping. $\sigma^{bal}_{l}$ drops as damping increases. Inset shows $\sigma^{bal}_{l}$ at 100K, 200K and 300K. As the bottom surface changes from free-standing to clamped ($\Xi$ = 0.1), $\sigma^{bal}_{l}$ reduces by $\sim$12\%,  $\sim$12.6\% and  $\sim$13.1\% at 100K, 200K and 300K, respectively . }
\label{fig:wire_kbal}
\end{figure}

The ballistic $\sigma_{l}$ for square SiNWs is calculated using their phonon dispersions. The conductance is calculated using Eq. (\ref{eq:kbal_1D}) assuming semi-infinite extensions along the wire growth axis (X-axis) and CBC on the periphery (Y and Z axis) of the wire (Fig.\ref{fig:nw_unitcell}) . Clamping the bottom surface affects the $\sigma^{bal}_{l}$ stronger at higher temperature compared to the lower temperature (Fig. \ref{fig:wire_kbal}a). Figure \ref{fig:wire_kbal}b shows the $\sigma^{bal}_{l}$ at 300K. The reduction in $\sigma^{bal}_{l}$ from free-standing wire to a clamped wire ($\Xi$ = 0.1) is $\sim$13.1\%. Hence, fixing the surface atoms have a strong impact of the lattice thermal conductance in SiNW.

\section{Conclusions}
\label{sec:4}
The details for calculating the phonon dispersion in zinc-blende semiconductor structures using the modified Valence Force Field (MVFF) method have been outlined. The MVFF method has been applied to calculate the phonon spectra in confined nanowire structures with varying boundary conditions. The methodology and the computational requirements for the method has been provided. Comparison of original Keating VFF with MVFF shows that, MVFF provides accurate phonon dispersion but at the expense of higher computational demands. Different VFF models can be used for obtaining the solution for physical quantities depending on the type of application, size of the structure and the computational resources available. We believe that the MVFF model will provide better phonon dispersion in ultra-scaled nanostructures. The MVFF method will be crucial in understanding and modeling the thermal properties of ultra-scaled semiconductor devices.

%
\section*{Acknowledgments}
The authors would like to acknowledge the computational resources from nanoHUB.org, an National Science Foundation (NSF) funded, NCN project. Financial support from MSD Focus Center, one of six research centers funded under the Focus Center Research Program (FCRP), a Semiconductor Research Corporation (SRC) entity and by the Nanoelectronics Research Initiative (NRI) through the Midwest Institute for Nanoelectronics Discovery (MIND) are also acknowledged.

\section*{APPENDICES}

\appendix
\numberwithin{equation}{section}
\section{Details of bulk zinc-blende coplanar interaction}
\label{app:bulk_int}

The coplanar interactions are important to obtain the flat nature of the acoustic phonon branches in Si and Ge \cite{VFF_mod_herman}. There are 21 such interactions in a zinc-blende crystal. The normalized locations of all the atoms involved in the coplanar interactions are shown in Table \ref{tab:atom_coords}. The corresponding groups used for bulk phonon dispersion calculations are given in Table \ref{tab:cop_grp}.

\begin{table}[b!]
\centering
\caption{Normalized atomic coordinates ($[\overline{x},\overline{y},\overline{z}] = [x,y,z]/a_{0}$) used for coplanar interaction calculation. }
\label{tab:atom_coords}       
\begin{tabular}{|l|ccc||l|ccc|}
\hline
No. & $\overline{x}$ & $\overline{y}$ & $\overline{z}$ & No. & $\overline{x}$ & $\overline{y}$ & $\overline{z}$ \\
\hline 
  1* &    0 &        0 &        0 & 14 &     0   & 0.50   &-0.50 \\
  2* & 0.25 &   0.25 &   0.25 & 15 &-0.50  & -0.50   &      0 \\
  3 & 0.25 &   -0.25 &   -0.25 & 16 & -0.50 &        0  &  0.50\\
  4 &-0.25  &  0.25 &  -0.25 & 17 &      0  & -0.50  &  0.50\\
  5 & -0.25 &  -0.25 &   0.25 &  18 & 0.25  &  0.75  &  0.75\\
  6  &     0  &  0.50  &  0.50 & 19 & -0.25 &   0.75 &   0.25\\
  7 & 0.50  &       0   & 0.50 & 20 & -0.25 &   0.25 &   0.75\\
  8 & 0.50  &  0.50   &      0 &   21 & 0.75  &  0.25 &   0.75\\
  9 &      0  & -0.50  & -0.50 & 22 & 0.75  & -0.25 &   0.25\\
  10 & 0.50  & -0.50   &      0 & 23 & 0.25  & -0.25 &   0.75\\
  11 & 0.50   &      0  & -0.50 &   24 & 0.75  &  0.75 &   0.25\\
  12 &-0.50  &       0  & -0.50 &   25 & 0.75  &  0.25 &  -0.25\\
  13 &-0.50  &  0.50   &      0 & 26 & 0.25  &  0.75 &  -0.25 \\
\hline
\multicolumn{8}{l}{*Belong to the main bulk unitcell used for DM calculation.}\\
\end{tabular}
\end{table}

\begin{table}[t!]
\centering
\caption{Atoms forming the coplanar interaction groups. 4 atoms in each group.}
\label{tab:cop_grp}       
\begin{tabular}{|l|cccc||l|cccc|}
\hline
No. & \multicolumn{4}{|c|}{Members} & No. & \multicolumn{4}{|c|}{Members} \\
\hline
 1	&  2   &  1   &  3  &   9  & 12 &  1  &   2  &    6  &  18  \\
 2  &  2   &  1   &  4  &  12  & 13 & 1   &  2   &  7  &  21  \\
 3  &  2   &  1   &  5  &  15  &  14 & 1   &  2  &   8  &  24  \\ 
 4  &  3   &  1   &  2   &  6  &  15 & 6   &  2   &  7  &   22	  \\
 5  &  3   &  1   &  4  &  13  &  16 & 6   &  2  &   8  &  25    \\
 6  &  3   &  1   &  5   & 16  &  17 & 7   &  2   &  6  &  19  \\
 7  &  4   &   1  &   2  &   7  &  18 & 7  &   2  &   8  &   26  \\
 8  &  4   &   1  &   3  &  10  &  19 & 8  &   2  &   6  &  20  \\
 9  &  4   &   1  &   5  &  17  &  20 & 8  &   2  &   7  &  23   \\
 10  &  5  &   1  &   2  &   8 &   21 &   5  &   1  &   4  &  14    \\
 11 &  5   &   1  &   3  &  11 &      &      &      &      &       \\
\hline
\multicolumn{10}{l}{*Atom numbers are same as shown in Table \ref{tab:atom_coords}.}\\
\end{tabular}
\end{table}

\section{Derivation of dynamical matrix from the equation of motion}
\label{app:dynmat_derive}
A crystal in equilibrium has zero total force. However, in the presence of perturbations like lattice vibrations, etc. a small restoring force works on the system. The total force ($F_{total}$) under small perturbation is given by the Taylor series expansion as,
\begin{eqnarray}
\label{eq:total_force}
F_{total} & = & -\sum_{i \in N} \frac{\partial U}{\partial \Delta R_{i}} \,(=\,0 \,at\,eqb.) \nonumber \\
          &  & - \frac{1}{2}\sum_{i,j \in N} \frac{\partial^2 U}{\partial \Delta R_{i} \partial \Delta R_{j}} \cdot \Delta R_{j} + \ldots
\end{eqnarray}
where, N represents all the atoms present in the system and U is the potential energy of the system. In Eq. (\ref{eq:total_force}) first term in RHS is zero under equilibrium. The next non-zero term is the second term in Eq. (\ref{eq:total_force}). Under harmonic approximation, only the second term is considered and the higher order (anharmonic) terms are neglected. Now combining Eq. (\ref{eq:motion}) and Eq. (\ref{eq:total_force}) one can obtain the following,
\begin{eqnarray}
\label{eq:dm_derive}
F_{total} & = & \sum_{i \in N} m_{i}\frac{\partial^2}{\partial t^2} \Delta R_{i}  \nonumber \\
          & = & -\frac{1}{2}\sum_{i,j \in N} \frac{\partial^2 U}{\partial \Delta R_{i} \partial \Delta R_{j}} \cdot \Delta  R_{j} \\
          & = & DR
\end{eqnarray} 
where, D is called the `Dynamical matrix' and R is a column vector of displacement for each atom given as,

\begin{equation}
\label{eq:DM_expand}
D = 
\begin{bmatrix}
D(11) & D(12) & \cdots & D(1N) \\
D(21) & D(22) & \cdots & D(2N) \\
\vdots & \vdots &\ddots & \vdots  \\
D(N1) & D(N2) & \cdots & D(NN)
\end{bmatrix}
\end{equation}

\begin{equation}
\label{eq:DM_expand}
R^{T} = 
\begin{bmatrix}
\Delta R_{1} & \Delta R_{2} & \cdots & \Delta R_{N} 
\end{bmatrix}
\end{equation}
Definition of $D(ij)$ is given in Eq. (\ref{eq:Dij_def}).

\section{Treatment of surface atoms}
\label{app:A}
The damped displacement of the surface atom `j' can be represented the matrix $\Xi^{j}$ given as,

\begin{equation}
\label{eq:damp_def}
\Xi^{j} = 
\begin{bmatrix}
\epsilon^{j}_{x} & 0 & 0\\
0 & \epsilon^{j}_{y} & 0\\
0 & 0 & \epsilon^{j}_{z}
\end{bmatrix}
\end{equation}

Taking into account the individual components the displacement vector for the atom `j' we obtain,
\begin{equation}
\tilde{r}^{j}_{n} = \epsilon^{j}_{n} r^{j}_{n} \quad n \in [x,y,z] 
\end{equation}
This modifies eqn.\ref{eq:Dij_def} as,
\begin{equation}
\label{eq:new_dij}
\tilde{D}^{ij}_{mn} = \epsilon^{j}_{n} D^{ij}_{mn} 
\end{equation}
Combining Eq. (\ref{eq:damp_def}) and (\ref{eq:new_dij}) the dynamical matrix component between atom `i' and `j' can be represented as,

\begin{equation}
\label{eq:new_dij_expand}
\tilde{D}(ij) = 
\begin{bmatrix}
\epsilon^{i}_{x} D^{ij}_{xx} \epsilon^{j}_{x} & \epsilon^{i}_{x} D^{ij}_{xy}\epsilon^{j}_{y} & \epsilon^{i}_{x} D^{ij}_{xz} \epsilon^{j}_{z}\\
\epsilon^{i}_{y} D^{ij}_{yx} \epsilon^{j}_{x} & \epsilon^{i}_{y} D^{ij}_{yy} \epsilon^{j}_{y} & \epsilon^{i}_{y} D^{ij}_{yz} \epsilon^{j}_{z}\\
\epsilon^{i}_{z} D^{ij}_{zx} \epsilon^{j}_{x} & \epsilon^{i}_{z} D^{ij}_{zy} \epsilon^{j}_{y} & \epsilon^{i}_{z} D^{ij}_{zz} \epsilon^{j}_{z}
 \end{bmatrix}
\end{equation}

which can be written in a compressed form as,

\begin{equation}
\label{eq:new_dij_compress}
\tilde{D}(ij) = \Xi^{i} D(ij) \Xi^{j}
\end{equation}

The value of $\epsilon_{x,y,z} \in [0,1] $, where completely free surface atoms have value 1 and completely tied atoms have value 0. 

\section{Inclusion of mass in the Dynamical matrix}
\label{app:B}
In Eq. (\ref{eq:DM_eig}) the mass of the atoms in on the RHS. It is convenient to include the mass in DM itself. This modifies the the LHS of the equation. The modified DM component between atom `i' and `j' thus, becomes,

\begin{equation}
\label{eq:new_dij_mass_red}
\tilde{D}(ij) =
\begin{bmatrix}
\frac{1}{\sqrt{m_i}} D^{ij}_{xx} \frac{1}{\sqrt{m_j}} & \frac{1}{\sqrt{m_i}} D^{ij}_{xy}\frac{1}{\sqrt{m_j}} & \frac{1}{\sqrt{m_i}} D^{ij}_{xz} \frac{1}{\sqrt{m_j}}\\
\frac{1}{\sqrt{m_i}} D^{ij}_{yx} \frac{1}{\sqrt{m_j}} & \frac{1}{\sqrt{m_i}} D^{ij}_{yy} \frac{1}{\sqrt{m_j}} & \frac{1}{\sqrt{m_i}} D^{ij}_{yz} \frac{1}{\sqrt{m_j}}\\
\frac{1}{\sqrt{m_i}} D^{ij}_{zx} \frac{1}{\sqrt{m_j}} & \frac{1}{\sqrt{m_i}} D^{ij}_{zy} \frac{1}{\sqrt{m_j}} & \frac{1}{\sqrt{m_i}} D^{ij}_{zz} \frac{1}{\sqrt{m_j}}
 \end{bmatrix}
\end{equation}
here $m_{i}$ and $m_{j}$ are the masses of atom `i' and `j' respectively. Eq.\ref{eq:new_dij_mass_red} can be written in a compressed manner as, 

\begin{equation}
\label{eq:new_dij_mass_comp}
\bar{D}(ij) = M^{-1}_{i} D(ij) M^{-1}_{j},
\end{equation}
where $M_{i}$ is given as,

\begin{equation}
\label{eq:mass_array}
M_{i} = 
\begin{bmatrix}
\sqrt{m_{i}} & 0 & 0\\
0 & \sqrt{m_{i}} & 0\\
0 & 0 & \sqrt{m_{i}}
\end{bmatrix}
\end{equation}

\section{Fitted analytical expressions for DM properties}
\label{app:DM_prop}
\begin{itemize}
\item
\textit{Atoms in a $\langle$100$\rangle$ SiNW unitcell}: The $N_{A}$ data obtained for the square wires till 6nm $\times$ 6nm can be fitted to a quadratic polynomial given as,
\begin{equation}
\label{NA_fitting}
N_{A}(W) = 27.57W^2 + 4.59W  
\end{equation}
Using Eq. (\ref{NA_fitting}) for a 16nm $\times$ 16nm SiNW gives around  7128 atoms. 

\item
\textit{Non-zero elements in a $\langle$100$\rangle$ SiNW DM}: The data for non-zero elements in the DM for SiNW with W till 6nm can be fitted to a quadratic polynomial given by,
\begin{equation}
\label{NZ_fitting}
NZ(W) =  3156W^2 -495.5W
\end{equation} 
Using Eq. (\ref{NZ_fitting}) for a 16nm $\times$ 16nm SiNW yields around 800117 non-zero elements in the DM.

\item
\textit{Relation of NZ elements to $N_A$ in a  $\langle$100$\rangle$ SiNW}: The number of non-zero (NZ) elements vary linearly with the number of atoms (Fig. \ref{NA_VS_NZ}). Fitting the NZ elements with $N_A$ for each wire under study following relations are obtained,
\begin{eqnarray}
\label{NZ_with_NA}
	NZ_{KVFF} & \approx & 109.3 \times N_{A} \\ 
	NZ_{MVFF} & \approx & 205.6 \times N_{A}  
\end{eqnarray}  
This shows that the number of NZ elements in MVFF method is roughly twice the NZ elements in KVFF model.

\begin{figure}[t!]
\centering
\includegraphics[width=3.0in,height=2.2in]{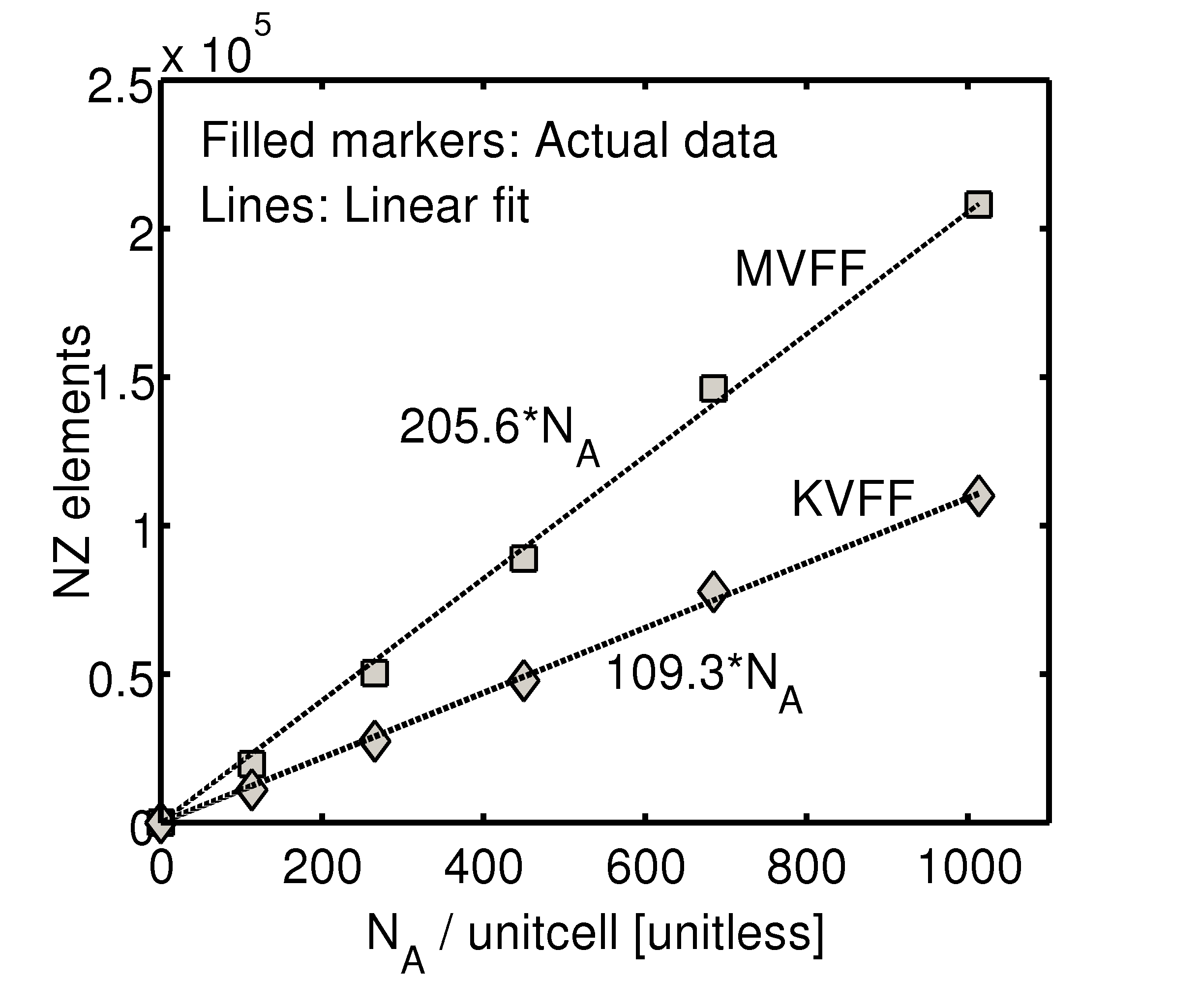} 
\caption{Variation in the number of non-zero (NZ) elements with $N_{A}$ for the two phonon models (1) Keating VFF and (2) Modified VFF. For both the cases NZ varies linearly with $N_{A}$. MVFF has roughly twice the number of NZ elements compared to KVFF model. }
\label{NA_VS_NZ}
\end{figure}

\item
\textit{Percentage fill factor for a $\langle$100$\rangle$ SiNW DM}: The percentage fill-factor can be derived using Eq. (\ref{NA_fitting}) and (\ref{NZ_fitting}) which leads to the following expression,
\begin{equation}
\label{percent_filled}
\%fill-factor(W) \approx  \frac{0.81}{W^2} \propto NA^{-1}
\end{equation} 
Eqn.\ref{percent_filled} estimates a 16nm $\times$ 16nm SiNW DM is filled only 0.32\%. 
This shows that the DM matrix is very sparsely filled for larger wires.

\item
\textit{Mean assembly time for $\langle$100$\rangle$ SiNW DM}: The data for mean $time_{asm}$ of the DM for SiNW with W till 6nm can be fitted to a quadratic polynomial given by,
\begin{equation}
\label{time_asm}
time_{asm}(W) =  0.9079W^2 -0.3789W \,secs
\end{equation} 
Using Eq. (\ref{time_asm}) a 16nm $\times$ 16nm SiNW DM is estimated to be assembled on single CPU in 226.35 secs.

\item
\textit{Eigen solution time per `k' point for $\langle$100$\rangle$ SiNW DM}: The eigen values are obtained using the `eig' solver in MATLAB \cite{eig_matlab}. The time needed for the solution of all the eigen values (with the eigen vectors) for each momentum vector `k' can be fitted to the following expression,
\begin{equation}
\label{eigen_time_fit}
time_{eigen} (W) = 5.4\times 10^{-3}W^{6.1}\;secs 
\end{equation}
Thus, the solution time goes with the sixth power of W. Also this expression gives an estimate time for the solution of one k point using the MATLAB eig solver (on a PC) as 1.19e5 seconds (~1.4 days). Thus, for very large systems parallel eigen solvers (like SCALAPACK \cite{slug}) as well as finding few eigen solutions (using eigs or other sparse eigen solvers, like LAPACK, etc.) is a feasible method.

\end{itemize}


\bibliographystyle{IEEEtran}      
\bibliography{jce_refs}   

\begin{thebibliography}{10}
\providecommand{\url}[1]{#1}
\csname url@samestyle\endcsname
\providecommand{\newblock}{\relax}
\providecommand{\bibinfo}[2]{#2}
\providecommand{\BIBentrySTDinterwordspacing}{\spaceskip=0pt\relax}
\providecommand{\BIBentryALTinterwordstretchfactor}{4}
\providecommand{\BIBentryALTinterwordspacing}{\spaceskip=\fontdimen2\font plus
\BIBentryALTinterwordstretchfactor\fontdimen3\font minus
  \fontdimen4\font\relax}
\providecommand{\BIBforeignlanguage}[2]{{%
\expandafter\ifx\csname l@#1\endcsname\relax
\typeout{** WARNING: IEEEtran.bst: No hyphenation pattern has been}%
\typeout{** loaded for the language `#1'. Using the pattern for}%
\typeout{** the default language instead.}%
\else
\language=\csname l@#1\endcsname
\fi
#2}}
\providecommand{\BIBdecl}{\relax}
\BIBdecl

\bibitem{JAP_anantram}
A.~Buin, A.~Verma, and M.~Anantram, ``{Carrier-phonon interaction in small
  cross-sectional silicon nanowires},'' \emph{Journal of Applid Physics}, vol.
  104, no. 053716, 2008.

\bibitem{buin_sinw_phonon}
\BIBentryALTinterwordspacing
A.~K. Buin, A.~Verma, A.~Svizhenko, and M.~P. Anantram, ``{Significant
  Enhancement of Hole Mobility in [110] Silicon Nanowires Compared to Electrons
  and Bulk Silicon},'' \emph{Nano Letters}, vol.~8, no.~2, pp. 760--765, 2008,
  pMID: 18205425. [Online]. Available:
  \url{http://pubs.acs.org/doi/abs/10.1021/nl0727314}
\BIBentrySTDinterwordspacing

\bibitem{Mingo_kappa}
N.~Mingo and L.~Yang, ``{Phonon transport in nanowires coated with an amorphous
  material: An atomistic Green's function approach},'' \emph{Phys. Rev. B},
  vol.~68, no.~24, p. 245406, Dec 2003.

\bibitem{mingo_ph}
N.~Mingo, L.~Yang, D.~Li, and A.~Majumdar, ``{Predicting the Thermal
  Conductivity of Si and Ge Nanowires},'' \emph{Nano Letters}, vol.~3, no.~12,
  pp. 1713--1716, 2003.

\bibitem{thermal_cond_dim}
J.~Wang and J.-S. Wang, ``{Dimensional crossover of thermal conductance in
  nanowires},'' \emph{Applied Physics Letters}, vol.~90, no.~24, p. 241908,
  2007.

\bibitem{SINW_110_phonon}
H.~Peelaers, B.~Partoens, and F.~M. Peeters, ``{Phonon Band Structure of Si
  Nanowires: A Stability Analysis},'' \emph{Nano Letters}, vol.~9, no.~1, pp.
  107--111, 2009.

\bibitem{Keating_VFF}
P.~N. Keating, ``{Effect of Invariance Requirements on the Elastic Strain
  Energy of Crystals with Application to the Diamond Structure},'' \emph{Phys.
  Rev.}, vol. 145, no.~2, pp. 637--645, 1966.

\bibitem{VFF_mod_herman}
Z.~Sui and I.~P. Herman, ``{Effect of strain on phonons in {Si, Ge, and Si/Ge}
  heterostructures},'' \emph{Phys. Rev. B}, vol.~48, no.~24, pp.
  17\,938--17\,953, 1993.

\bibitem{VFF_mod_zunger}
H.~Fu, V.~Ozolins, and Z.~Alex, ``Phonons in {GaP} quantum dots,'' \emph{Phys.
  Rev. B}, vol.~59, no.~4, pp. 2881--2887, 1999.

\bibitem{McMurry_VFF}
H.~McMurry, A.~Solbrig~Jr., and J.~Boyter, ``{The use of valence force
  potentials in calculating crystal vibrations},'' \emph{Journal of Physics and
  Chemistry of Solids}, vol.~28, no.~12, pp. 2359 -- 2368, 1967.

\bibitem{bcm_weber}
W.~Weber, ``{Adiabatic bond charge model for the phonons in diamond, {Si, Ge,
  and $\alpha{}$-Sn}},'' \emph{Phys. Rev. B}, vol.~15, no.~10, pp. 4789--4803,
  May 1977.

\bibitem{BCM_model}
K.~Rustagi and W.~Weber, ``{Adiabatic Bond charge model for the phonons in
  {A(III)B(V)} semiconductors},'' \emph{Solid State Communications}, vol.~18,
  pp. 673--675, 1976.

\bibitem{jauho_method}
T.~Markussen, A.-P. Jauho, and M.~Brandbyge, ``{Heat Conductance Is Strongly
  Anisotropic for Pristine Silicon Nanowires},'' \emph{Nano Letters}, vol.~8,
  no.~11, pp. 3771--3775, 2008.

\bibitem{six_param_VFF}
H.~L. McMurry, A.~W. Solbrig, B.~J. K., and C.~Noble, ``The use of valence
  force potentials in calculating crystal vibrations,'' \emph{J. Phys. Chem.
  Solids}, vol.~28, pp. 2359--2368, 1967.

\bibitem{continuum_model}
J.~Zou and A.~Balandin, ``Phonon heat conduction in a semiconductor nanowire,''
  \emph{Journal of Applied Physics}, vol.~89, no.~5, pp. 2932--2938, 2001.

\bibitem{sinw_cv}
Y.~Zhang, J.~X. Cao, Y.~Xiao, and X.~H. Yan, ``Phonon spectrum and specific
  heat of silicon nanowires,'' \emph{Journal of Applied Physics}, vol. 102,
  no.~10, p. 104303, 2007.

\bibitem{strain_effect_1}
X.~Li, K.~Maute, M.~L. Dunn, and R.~Yang, ``Strain effects on the thermal
  conductivity of nanostructures,'' \emph{Phys. Rev. B}, vol.~81, no.~24, p.
  245318, Jun 2010.

\bibitem{SINW_111}
T.~Thonhauser and G.~D. Mahan, ``{Phonon modes in Si [111] nanowires},''
  \emph{Phys. Rev. B}, vol.~69, no.~7, p. 075213, Feb 2004.

\bibitem{strain_effect_2}
H.~Zhao, Z.~Tang, G.~Li, and N.~R. Aluru, ``Quasiharmonic models for the
  calculation of thermodynamic properties of crystalline silicon under
  strain,'' \emph{Journal of Applied Physics}, vol.~99, no.~6, p. 064314, 2006.

\bibitem{anharmonic}
O.~L. Lazarenkova, P.~von Allmen, F.~Oyafuso, S.~Lee, and G.~Klimeck, ``{Effect
  of anharmonicity of the strain energy on band offsets in semiconductor
  nanostructures},'' \emph{Applied Physics Letters}, vol.~85, no.~18, pp.
  4193--4195, 2004.

\bibitem{DM_element_reduction}
Z.~W. Hendrikse, M.~O. Elout, and W.~J.~A. Maaskant, ``{Computation of the
  independent elements of the dynamical matrix},'' \emph{Computer Physics
  Communications}, vol.~86, no.~3, pp. 297 -- 311, 1995.

\bibitem{Land}
R.~Landauer, ``{Spatial variation of currents and fields due to localized
  scatterers in metallic conduction},'' \emph{IBM J. Res. Dev.}, vol.~1, no.~3,
  pp. 223--231, 1957.

\bibitem{Wallace}
D.~C. Wallace, ``{Thermodynamics of Crystals},'' \emph{Dover Publications,
  Mineola New York}, 1998.

\bibitem{gparam_1}
B.~A. Weinstein and G.~J. Piermarini, ``{Raman scattering and phonon dispersion
  in Si and GaP at very high pressure},'' \emph{Phys. Rev. B}, vol.~12, no.~4,
  pp. 1172--1186, Aug 1975.

\bibitem{CRS_CCS}
\BIBentryALTinterwordspacing
J.~Dongarra, ``{Survey of Sparse Matrix Storage Formats},'' 1995. [Online].
  Available: \url{http://www.netlib.org/linalg/html_templates/node90.html}
\BIBentrySTDinterwordspacing

\bibitem{eig_matlab}
\BIBentryALTinterwordspacing
Mathworks, ``{Matlab eig reference},'' 2010. [Online]. Available:
  \url{http://www.mathworks.com/help/techdoc/ref/eig.html}
\BIBentrySTDinterwordspacing

\bibitem{eigs_matlab}
\BIBentryALTinterwordspacing
------, ``{Matlab eig reference},'' 2010. [Online]. Available:
  \url{http://www.mathworks.com/help/techdoc/ref/eigs.html}
\BIBentrySTDinterwordspacing

\bibitem{nemo3d}
G.~Klimeck, F.~Oyafuso, T.~B. Boykin, R.~C. Bowen, and P.~von Allmen,
  ``{Development of a Nanoelectronic 3-D (NEMO 3-D) Simulator for Multimillion
  Atom Simulations and Its Application to Alloyed Quantum Dots},''
  \emph{Computer Modeling in Engineering and Science (CMES)}, vol.~3, no.~5,
  pp. 601--642, 2002.

\bibitem{bslab}
\BIBentryALTinterwordspacing
A.~Paul, M.~Luisier, N.~Neophytou, R.~Kim, J.~Geng, M.~McLennan, M.~Lundstrom,
  and G.~Klimeck, ``{Band Structure Lab},'' May 2006. [Online]. Available:
  \url{http://nanohub.org/resources/1308}
\BIBentrySTDinterwordspacing

\bibitem{bulk_si_exp_phon}
G.~Nilsson and G.~Nelin, ``{Study of the Homology between Silicon and Germanium
  by Thermal Neutron Spectrometry},'' \emph{Phys. Rev. B}, vol.~6, no.~10, pp.
  3777--3786, 1972.

\bibitem{ioffe_online}
``{Electronic archive, New Semiconductor Materials - Characteristics and
  Properties},'' Ioffe Physico-Technical Institute Website, 2001,
  http://www.ioffe.ru/SVA/NSM/Semicond/.

\bibitem{Madelung}
O.~Madelung, ``{Semiconductors - HandBook},'' \emph{Springer 3rd ed.}, 2004.

\bibitem{gparam_2}
S.~de~Gironcoli, ``{Phonons in Si-Ge systems: An ab initio
  interatomic-force-constant approach},'' \emph{Phys. Rev. B}, vol.~46, no.~4,
  pp. 2412--2419, Jul 1992.

\bibitem{gparam_3}
R.~Eryi\ifmmode~\breve{g}\else \u{g}\fi{}it and I.~P. Herman, ``{Lattice
  properties of strained GaAs, Si, and Ge using a modified bond-charge
  model},'' \emph{Phys. Rev. B}, vol.~53, no.~12, pp. 7775--7784, Mar 1996.

\bibitem{olga_gparam}
O.~L. Lazarenkova, P.~von Allmen, F.~Oyafuso, S.~Lee, and G.~Klimeck, ``{An
  atomistic model for the simulation of acoustic phonons, strain distribution,
  and Grüneisen coefficients in zinc-blende semiconductors},''
  \emph{Superlattices and Microstructures}, vol.~34, no. 3-6, pp. 553 -- 556,
  2003.

\bibitem{slug}
L.~S. Blackford, J.~Choi, A.~Cleary, E.~D'Azevedo, J.~Demmel, I.~Dhillon,
  J.~Dongarra, S.~Hammarling, G.~Henry, A.~Petitet, K.~Stanley, D.~Walker, and
  R.~C. Whaley, \emph{{ScaLAPACK} Users' Guide}.\hskip 1em plus 0.5em minus
  0.4em\relax Philadelphia, PA: Society for Industrial and Applied Mathematics,
  1997.

\end{thebibliography}

\end{document}